\documentclass[lettersize,journal]{IEEEtran}
\usepackage{ifpdf}
\usepackage{cite}
\usepackage[pdftex]{graphicx}

\usepackage[T1]{fontenc}

\usepackage{afterpage}
\usepackage{bm}
\usepackage{listings}
\usepackage{xcolor}
\usepackage{graphicx}
\usepackage{mathtools}
\usepackage{float}
\usepackage{multirow}
\usepackage[font=small]{caption}
\usepackage[font=footnotesize]{subcaption}

\usepackage{placeins}
\usepackage{url}
\usepackage{epstopdf}
\usepackage{tikz}
\usepackage{pgfplots}
\pgfplotsset{compat=1.15} 

\newlength\fheight
\newlength\fwidth

\usetikzlibrary{backgrounds,patterns,calc}
\usepackage{booktabs}
\usepackage{footnote}
\makesavenoteenv{table}
\makesavenoteenv{tabular}
\usepackage{flushend}
\usepackage{amsmath}
\usepackage{amsthm}
\usepackage{amssymb}
\usepackage{tabularx}
\usepackage{booktabs}
\usepackage{placeins}
\usepackage{graphicx}
\usepackage{color}
\usepackage{colortbl}
\usepackage{bbold}
\DeclareMathOperator*{\argmax}{arg\,max}
\usepackage[acronyms,nonumberlist,nopostdot,nomain,nogroupskip]{glossaries}
\usepackage{hyperref}
\usepackage{tabularx}
\usepackage[linesnumbered,ruled,vlined]{algorithm2e}
\usepackage{algpseudocode}
\newsavebox{\bigimage}
\usepackage{setspace}

\usepackage{tcolorbox}
\tcbuselibrary{theorems}

\newtcbtheorem[no counter]{mytheo}{Summary}%
{colback=white,colframe=black,fonttitle=\bfseries}{th}

\makeatother

\usepackage{glossaries}

\newacronym{3gpp}{3GPP}{3rd Generation Partnership Project}
\newacronym{adc}{ADC}{Analog to Digital Converter}
\newacronym{5g}{5G}{5th generation}
\newacronym{6g}{6G}{6th generation}
\newacronym{aimd}{AIMD}{Additive Increase Multiplicative Decrease}
\newacronym{am}{AM}{Acknowledged Mode}
\newacronym{amc}{AMC}{Adaptive Modulation and Coding}
\newacronym{aqm}{AQM}{Active Queue Management}
\newacronym{awgn}{AWGN}{Additive White Gaussian Noise}
\newacronym{balia}{BALIA}{Balanced Link Adaptation}
\newacronym{bdp}{BDP}{Bandwidth-Delay Product}
\newacronym{qos}{QoS}{Quality of Service}
\newacronym{pqos}{PQoS}{predictive quality of service}
\newacronym{bf}{BF}{Beamforming}
\newacronym{cc}{CC}{Congestion Control}
\newacronym{cdf}{CDF}{Cumulative Distribution Function}
\newacronym{pdf}{pdf}{probability density function}
\newacronym{cn}{CN}{Core Network}
\newacronym{cqi}{CQI}{Channel Quality Information}
\newacronym{cp}{CP}{Control Plane}
\newacronym{csirs}{CSI-RS}{Channel State Information - Reference Signal}
\newacronym{dc}{DC}{Dual Connectivity}
\newacronym{dce}{DCE}{Direct Code Execution}
\newacronym{dci}{DCI}{Downlink Control Information}
\newacronym{dl}{DL}{downlink}
\newacronym{dmr}{DMR}{Deadline Miss Ratio}
\newacronym{dmrs}{DMRS}{DeModulation Reference Signal}
\newacronym{e2e}{E2E}{end-to-end}
\newacronym{ecn}{ECN}{Explicit Congestion Notification}
\newacronym{edf}{EDF}{Earliest Deadline First}
\newacronym{enb}{eNB}{evolved Node Base}
\newacronym{epc}{EPC}{Evolved Packet Core}
\newacronym{es}{ES}{Edge Server}
\newacronym{fdma}{FDMA}{Frequency Division Multiple Access}
\newacronym{fdd}{FDD}{Frequency Division Duplexing}
\newacronym[firstplural=Radio Access Technologies (RATs)]{rat}{RAT}{Radio Access Technology}
\newacronym{fs}{FS}{Fast Switching}
\newacronym{ftp}{FTP}{File Transfer Protocol}
\newacronym{gnb}{gNB}{Next Generation Node~B}
\newacronym{harq}{HARQ}{Hybrid Automatic Repeat reQuest}
\newacronym{hetnet}{HetNet}{Heterogeneous Network}
\newacronym{hh}{HH}{Hard Handover}
\newacronym{hol}{HOL}{Head-of-Line}
\newacronym{ia}{IA}{Initial Access}
\newacronym{ieee}{IEEE}{Institute of Electrical and Electronics Engineers}
\newacronym{imt}{IMT}{International Mobile Telecommunication}
\newacronym{iot}{IoT}{Internet of Things}
\newacronym{ldpc}{LDPC}{Low-Density Parity Check}
\newacronym{los}{LOS}{Line of Sight}
\newacronym{lte}{LTE}{Long Term Evolution}
\newacronym{m2m}{M2M}{Machine to Machine}
\newacronym{ml}{ML}{machine learning}
\newacronym{mac}{MAC}{Medium Access Control}
\newacronym{mc}{MC}{Multi-Connectivity}
\newacronym{mcs}{MCS}{Modulation and Coding Scheme}
\newacronym{mec}{MEC}{Mobile Edge Cloud}
\newacronym{mi}{MI}{Mutual Information}
\newacronym{mimo}{MIMO}{Multiple Input, Multiple Output}
\newacronym{mmwave}{mmWave}{millimeter wave}
\newacronym{mptcp}{MPTCP}{Multipath TCP}
\newacronym{mr}{MR}{Maximum Rate}
\newacronym{mss}{MSS}{Maximum Segment Size}
\newacronym{mtd}{MTD}{Machine-Type Device}
\newacronym{mtu}{MTU}{Maximum Transmission Unit}
\newacronym{nfv}{NFV}{Network Function Virtualization}
\newacronym{nlos}{NLOS}{Non-Line-of-Sight}
\newacronym{nlosv}{NLOSv}{Vehicle Non-Line-of-Sight}
\newacronym{nr}{NR}{New Radio}
\newacronym{ofdm}{OFDM}{Orthogonal Frequency Division Multiplexing}
\newacronym{pdcch}{PDCCH}{Physical Downlink Control Channel}
\newacronym{pdcp}{PDCP}{Packet Data Convergence Protocol}
\newacronym{pdsch}{PDSCH}{Physical Downlink Shared Channel}
\newacronym{pdu}{PDU}{Packet Data Unit}
\newacronym{pf}{PF}{Proportional Fair}
\newacronym{pgw}{PGW}{Packet Gateway}
\newacronym{phy}{PHY}{Physical}
\newacronym{pbch}{PBCH}{Physical Broadcast Channel}
\newacronym[plural=\gls{mme}s,firstplural=Mobility Management Entities (MMEs)]{mme}{MME}{Mobility Management Entity}
\newacronym{prb}{PRB}{Physical Resource Block}
\newacronym{pss}{PSS}{Primary Synchronization Signal}
\newacronym{pscch}{PSCCH}{Physical Sidelink Control Channel}
\newacronym{pucch}{PUCCH}{Physical Uplink Control Channel}
\newacronym{pusch}{PUSCH}{Physical Uplink Shared Channel}
\newacronym{rach}{RACH}{Random Access Channel}
\newacronym{ran}{RAN}{Radio Access Network}
\newacronym{red}{RED}{Random Early Detection}
\newacronym{rf}{RF}{Radio Frequency}
\newacronym{rlc}{RLC}{Radio Link Control}
\newacronym{rlf}{RLF}{Radio Link Failure}
\newacronym{rrc}{RRC}{Radio Resource Control}
\newacronym{rrm}{RRM}{Radio Resource Management}
\newacronym{rr}{RR}{Round Robin}
\newacronym{rs}{RS}{Remote Server}
\newacronym{rsrp}{RSRP}{Reference Signal Received Power}
\newacronym{rss}{RSS}{Received Signal Strength}
\newacronym{rtt}{RTT}{Round Trip Time}
\newacronym{rw}{RW}{Receive Window}
\newacronym{rx}{RX}{Receiver}
\newacronym{sa}{SA}{standalone}
\newacronym{sack}{SACK}{Selective Acknowledgment}
\newacronym{sap}{SAP}{Service Access Point}
\newacronym{sc}{SC}{Single Carrier}
\newacronym{sch}{SCH}{Secondary Cell Handover}
\newacronym{scoot}{SCOOT}{Split Cycle Offset Optimization Technique}
\newacronym{sdma}{SDMA}{Spatial Division Multiple Access}
\newacronym{sinr}{SINR}{Signal to Interference plus Noise Ratio}
\newacronym{sl}{SL}{Sidelink}
\newacronym{sm}{SM}{Saturation Mode}
\newacronym{snr}{SNR}{Signal-to-Noise-Ratio}
\newacronym{son}{SON}{Self-Organizing Network}
\newacronym{ss}{SS}{Synchronization Signal}
\newacronym{srs}{SRS}{Sounding Reference Signal}
\newacronym{sss}{SSS}{Secondary Synchronization Signal}
\newacronym{tb}{TB}{Transport Block}
\newacronym{tcp}{TCP}{Transmission Control Protocol}
\newacronym{tdd}{TDD}{Time Division Duplexing}
\newacronym{tdma}{TDMA}{Time Division Multiple Access}
\newacronym{tfl}{TfL}{Transport for London}
\newacronym{tm}{TM}{Transparent Mode}
\newacronym{trp}{TRP}{Transmitter Receiver Pair}
\newacronym{tti}{TTI}{Transmission Time Interval}
\newacronym{ttt}{TTT}{Time-to-Trigger}
\newacronym{tx}{TX}{Transmitter}
\newacronym{ue}{UE}{User Equipment}
\newacronym{ul}{UL}{uplink}
\newacronym{uml}{UML}{Unified Modeling Language}
\newacronym{um}{UM}{Unacknowledged Mode}
\newacronym{utc}{UTC}{Urban Traffic Control}
\newacronym{vm}{VM}{Virtual Machine}
\newacronym{rsrq}{RSRQ}{Reference Signal Received Quality}
\newacronym{rssi}{RSSI}{Received Signal Strength Indicator}
\newacronym{crs}{CRS}{Cell Reference Signal}
\newacronym{nsa}{NSA}{Non Stand Alone}
\newacronym{mrdc}{MR-DC}{Multi \gls{rat} \gls{dc}}
\newacronym{endc}{EN-DC}{E-UTRAN-\gls{nr} \gls{dc}}
\newacronym{5gc}{5GC}{5G Core}
\newacronym{si}{SI}{Study Item}
\newacronym{iab}{IAB}{Integrated Access and Backhaul}
\newacronym{wf}{WF}{Wired-first}
\newacronym{hqf}{HQF}{Highest-quality-first}
\newacronym{pa}{PA}{Position-aware}
\newacronym{mlr}{MLR}{Maximum-local-rate}
\newacronym{wbf}{WBF}{Wired Bias Function}
\newacronym{mib}{MIB}{Master Information Block}
\newacronym{sib}{SIB}{Secondary Information Block}
\newacronym{rnti}{RNTI}{Radio Network Temporary Identifier}
\newacronym{dft}{DFT}{Discrete Fourier Transform}
\newacronym{kpi}{KPI}{Key Performance Indicator}
\newacronym{ppp}{PPP}{Poisson Point Process}
\newacronym{v2v}{V2V}{Vehicle-to-Vehicle}
\newacronym{wave}{WAVE}{Wireless Access in Vehicular Environments}
\newacronym{udp}{UDP}{User Datagram Protocol}
\newacronym{upa}{UPA}{Uniform Planar Array}
\newacronym{fec}{FEC}{Forward Error Correction}
\newacronym{v2x}{V2X}{Vehicle-To-Everything}
\newacronym{psfch}{PSFCH}{Physical Sidelink Feedback Channel}
\newacronym{pssch}{PSSCH}{Physical Sidelink Shared Channel}
\newacronym{csma}{CSMA}{Carrier Sense Multiple Access}
\newacronym{v2n}{V2N}{Vehicle-to-Network}
\newacronym{wlan}{WLAN}{Wireless Local Area Network}
\newacronym{cav}{CAV}{Connected and Autonomous Vehicle}
\newacronym{v2i}{V2I}{Vehicle-to-Infrastructure}
\newacronym{d2d}{D2D}{Device-to-Device}
\newacronym{c-its}{C-ITS}{Connected Intelligent Transportation System}
\newacronym{fr2}{FR2}{Frequency Range 2}
\newacronym{fr1}{FR1}{Frequency Range 1}
\newacronym{bs}{BS}{Base Station}
\newacronym{sdu}{SDU}{Service Data Unit}
\newacronym{csi}{CSI}{Channel State Information}
\newacronym{scs}{SCS}{Subcarrier Spacing}
\newacronym{sumo}{SUMO}{Simulation of Urban MObility}
\newacronym{prr}{PRR}{Packet Reception Ratio}
\newacronym{edca}{EDCA}{Enhanced Distribution Channel Access}
\newacronym{sdap}{SDAP}{Service Data Adaptation Protocol}
\newacronym{iiot}{IIoT}{Industrial Internet of Things}
\newacronym{agv}{AGV}{Automated Guided Vehicles}
\newacronym{cm}{C/M}{Controller/Master}
\newacronym{soa}{SoA}{State-of-the-Art}
\newacronym{snpn}{SNPN}{Standalone Non-Public Network}
\newacronym{pninpn}{PNI-NPN}{Public Network Interface Non-Public Network}
\newacronym{urllc}{URLLC}{Ultra-Reliable Low-Latency Communication}
\newacronym{embb}{eMBB}{enhanced Mobile BroadBand}
\newacronym{ai}{AI}{Artificial Intelligence}
\newacronym{mab}{MAB}{Multi-Armed Bandit}
\newacronym{su}{SU}{Scheduling Unit}
\newacronym{ra}{RA}{Random Agent}
\newacronym{na}{NA}{Neural Agent}
\newacronym{ucba}{UCB-A}{UCB Agent}
\newacronym{tsa}{TS-A}{Thompson Sampling Agent}
\newacronym{ucb}{UCB}{Upper Confidence Bound}
\newacronym{ts}{TS}{Thompson Sampling}
\newacronym{lts}{LTS}{Linear Thompson Sampling}
\newacronym{inf}{InF}{Indoor Factory}
\newacronym{infsl}{InF-SL}{Indoor Factory - Sparse Clutter, Low BS}
\newacronym{infdl}{InF-DL}{Indoor Factory - Dense Clutter, Low BS}
\newacronym{infsh}{InF-SH}{Indoor Factory - Sparse Clutter, High BS}
\newacronym{infdh}{InF-DH}{Indoor Factory - Dense Clutter, High BS}
\newacronym{us}{US}{Uplink Scheduler}
\newacronym{nn}{NN}{Neural Network}
\newacronym{das}{DAS}{Distributed Antenna System}
\newacronym{rb}{RB}{Resource Block}
\newacronym{rl}{RL}{Reinforcement Learning}
\newacronym{uav}{UAV}{Unmanned Aerial Vehicle}
\newacronym{5gacia}{5G-ACIA}{5G Alliance for Connected Industries and Automation}
\newacronym{drl}{DRL}{Deep Reinforcement Learning}
\newacronym{ru}{RU}{Resource Unit}
\newacronym{fci}{FCI}{Feedback Control Information}
\newacronym{ccmab}{CC-MAB}{Contextual Combinatorial Multi-Armed Bandit}
\newacronym{cmab}{CMAB}{Contextual Multi-Armed Bandit}
\newacronym{maccmab}{MA-CC-MAB}{Multi-Agent Contextual Combinatorial Multi-Armed Bandit}
\newacronym{cnlts}{C-NLTS}{combinatorial neural linear thompson sampling}
\newacronym{dnn}{DNN}{Deep Neural Network}
\newacronym{sgd}{SGD}{stochastic gradient descent}
\newacronym{dcnlts}{DC-NLTS}{Distributed Combinatorial Neural Linear Thompson Sampling}
\newacronym{nlts}{NLTS}{Neural Linear Thompson Sampling}
\newacronym{mdp}{MDP}{Markov Decision Process}
\newacronym{iid}{i.i.d.}{independent and identically distributed}
\newacronym{disnets}{DISNETS}{DIStributed combinatorial NEural linear Thompson Sampling}
\newacronym{gbs}{GBS}{grant-based scheduling}
\newacronym{sps}{SPS}{semi-persistent scheduling}
\newacronym{gfs}{GFS}{grant-free scheduling}
\usepackage{balance}
\begin{document}
	
	\title{
	A Distributed Neural Linear Thompson Sampling Framework to Achieve URLLC in Industrial IoT}
	
	\author{Francesco Pase,~\IEEEmembership{Student~Member,~IEEE}, Marco Giordani,~\IEEEmembership{Member,~IEEE}, Sara Cavallero,~\IEEEmembership{Student~Member,~IEEE}, Malte Schellmann, Josef Eichinger, Roberto Verdone,~\IEEEmembership{Senior Member,~IEEE}, Michele~Zorzi,~\IEEEmembership{Fellow,~IEEE}
		\thanks{Francesco Pase, Marco Giordani, and Michele Zorzi are with WiLab and the Department of Information Engineering (DEI) of the University of Padova, Italy. Email: \{pasefrance,giordani,zorzi\}@dei.unipd.it.
			
			Sara Cavallero, and Roberto Verdone are with WiLab and Dipartimento di Ingegneria dell'Energia Elettrica e dell'Informazione ``Guglielmo Marconi'' of the University of Bologna, Italy. Email: \{s.cavallero,roberto.verdone\}@unibo.it.
			
			Josef Eichinger and Malte Schellmann are with Huawei Technologies, Munich Research Center, Germany. Email: \{joseph.eichinger,malte.schellmann\}@huawei.com.
			
			This work has been carried out in the framework of the CNIT National Laboratory WiLab and the WiLab-Huawei Joint Innovation Center.
			This work was also partially supported by the European Union under the Italian
			National Recovery and Resilience Plan (NRRP) of NextGenerationEU,
			partnership on “Telecommunications of the Future” (PE0000001 - program
			“RESTART”).}}
	
	\markboth{Journal of \LaTeX\ Class Files,~Vol.~14, No.~8, August~2021}%
	{Shell \MakeLowercase{\textit{et al.}}: A Sample Article Using IEEEtran.cls for IEEE Journals}
	
	
	\maketitle
	
	\begin{abstract}
		Industrial Internet of Things (IIoT) networks will provide Ultra-Reliable Low-Latency Communication (URLLC) to support critical processes underlying the production chains. However, standard protocols for allocating wireless resources may not optimize the latency-reliability trade-off, especially for uplink communication. For example, centralized grant-based scheduling can ensure almost zero collisions, but introduces delays in the way resources are requested by the User Equipments (UEs) and granted by the gNB. In turn, distributed scheduling (e.g., based on random access), in which UEs autonomously choose the resources for transmission, may lead to potentially many collisions especially when the traffic increases. In this work we propose DIStributed combinatorial NEural linear Thompson Sampling (DISNETS), a novel scheduling framework that combines the best of the two worlds. By leveraging a feedback signal from the gNB and reinforcement learning, the UEs are trained to autonomously optimize their uplink transmissions by selecting the available resources to minimize the number of collisions, without additional message exchange to/from the gNB. DISNETS is a distributed, multi-agent adaptation of the Neural Linear Thompson Sampling (NLTS) algorithm, which has been further extended to admit multiple parallel actions. We demonstrate the superior performance of DISNETS in addressing URLLC in IIoT scenarios compared to other baselines.
	\end{abstract}
	
		\begin{tikzpicture}[remember picture,overlay]
		\node[anchor=north,yshift=-10pt] at (current page.north) {\parbox{\dimexpr\textwidth-\fboxsep-\fboxrule\relax}{
				\centering\footnotesize This paper has been submitted to the IEEE Transaction on Wireless Communications. Copyright may change without notice.
		}};
	\end{tikzpicture}
	\glsresetall
	
	\section{Introduction}
	\label{sec:introduction}
	
	As \gls{5g} systems are already in the full implementation phase, the research community is now discussing future \gls{6g} networks and their requirements~\cite{giordani2020toward}. One of the driving forces in \gls{6g} is the design of new communication interfaces and architectures for \gls{iiot} networks, in which sensors, wearables, actuators, and robots are wirelessly interconnected in factories to enable analytics, diagnostics, monitoring, asset tracking, as well as process, regulatory, supervisory, and safety applications~\cite{wollschlaeger2017future,cheng2018industrial,vitturi2019industrial}. In this scenario, IIoT poses strict communication requirements to achieve almost real-time coordination, control, and sensing~\cite{lee2015cyber}. Specifically, these requirements translate into latency (less than $1$ ms in the radio
	part) and reliability (up to $99.99999\%$) constraints, thus calling for \gls{urllc}~\cite{5gacia,22104}. 
	
	As far as latency is concerned, the time introduced by \gls{ran} operations, from routing and resource allocation to modulation, represents one of the most significant sources of delay. In particular, the \gls{cm}, i.e., the \gls{gnb} of the \gls{iiot} network, should be able to provide almost immediate channel access to \glspl{ue} in the factory floor, ideally as soon as they have data to send. This is especially critical for uplink communication, as additional energy and computational constraints at the end nodes may further delay the time it takes to access the channel~\cite{oueis2016uplink,patriciello2019impact}.
	
	
	However, the \gls{5g} standard is unlikely to provide resource allocation in short time, mainly due to the intrinsic limitations of current channel access schemes~\cite{Le2019,cuozzo2022enabling}. 
	Notably, the \gls{3gpp} \gls{nr} specifications for 5G networks~\cite{38300} support three options to allocate uplink resources: {\gls{gbs}}, {\gls{sps}},\footnote{Formally, 3GPP 5G NR specifications define \gls{sps} for downlink scheduling, and Configured Grant (CG) Type 1 and Type 2 for uplink scheduling, where Type 2 is similar to SPS with minor modifications~\cite{larranaga20225g}. For simplicity, in the remainder of this paper we will refer to \gls{sps} also for uplink scheduling, even though our implementation is based on CG Type 2 specifications.} and {\gls{gfs}}~\cite{lin20195g}. \Gls{gbs}~\cite{36213} is fully centralized, and requires: (i) the \glspl{ue} to use the \gls{pucch} to ask the uplink scheduler for being scheduled; (ii) the gNB to communicate via the \gls{pdcch} to the \glspl{ue} which resources can be used for transmission; (iii) the \glspl{ue} to transmit their data blocks through the \gls{pusch}; and (iv) the gNB to provide the communication acknowledgment via \gls{harq}. This procedure requires at least two \glspl{rtt} from when data arrives in the buffer until it can be properly scheduled, which may prohibitively increase the communication delay.
	\Gls{sps}~\cite{R1-167309} is also fully centralized, but permits the \gls{gnb} to pre-allocate radio resources 
	without explicit scheduling requests and grants from/to the \glspl{ue}, 
	thus reducing the latency.
	However, the periodicity of scheduling grants is defined by \gls{rrc} signaling at the session establishment based on the predicted traffic at the \glspl{ue}, and may cause large and systematic delays in case of errors in those predictions. 
	On the other extreme, \gls{gfs}~\cite{liu2020analyzing} is fully distributed, and the \glspl{ue} autonomously choose radio resources to be used for transmission, thereby eliminating the need to wait for scheduling grants. On the downside, uncoordinated resource allocation may lead to collisions, and trigger re-transmissions accordingly, which again pose additional latency concerns.
	
	In this context, \gls{ml} has emerged as a promising tool to optimize network performance, including minimizing latency during resource allocation. 
	Still, most of the literature focuses on centralized and downlink algorithms, e.g.,~\cite{kasgari:2019:icc_model_free}, which however are not scalable as the density of the network increases.
	In the area of distributed learning, \gls{mab} algorithms~\cite{intro_mab}, and especially \gls{lts}~\cite{tutorial_thompson_sampling}, gained popularity to address the problem of resource allocation. 
	However, these schemes are often too simple to model complex network dynamics, and work with the assumption of linear dependency of data~\cite{globecom_mab}.
	A promising attempt to overcome this limitation was made with the \gls{nlts} algorithm~\cite{Riquelme2018}, which still assumes that the \gls{ml} agent can only play single actions, i.e., UEs transmit through one single orthogonal channel, which may increase the latency beyond \gls{urllc}~requirements.

	\subsection{Contributions}
	
	To solve these issues, in this work we propose a new distributed framework for resource allocation called \gls{disnets}, which is built upon two cardinal principles. First, it consists of a UE-centric architecture in which resource allocation decisions are made by the local UEs, ``disaggregated'' from the network, and without pre-defined scheduling requests and/or grants. Second, UEs rely on \gls{ml} to optimize resource allocation, which allows to minimize the probability of collisions and reduce the latency due to re-transmissions.
	To this aim, 
	our contributions are the following:
	\begin{itemize}
		\item We formalize the problem of distributed resource allocation as a \gls{maccmab} problem, where \glspl{ue} autonomously choose the physical resources to use for transmission. The problem is solved using \gls{disnets}, built on top of the \gls{nlts} algorithm~\cite{Riquelme2018}, which combines \gls{dnn} and \gls{lts} to optimize network operations. Specifically,  the original \gls{nlts} implementation is extended into the proposed \gls{disnets} solution by allowing agents to take more than one action at each scheduling opportunity, i.e., using multiple orthogonal channels in parallel in the same scheduling opportunity, which is important to provide URLLC. 
		\item We propose the design and structure of a new control signaling scheme, referred to as \gls{fci}, and used by the UEs to train and learn how to allocate resources using \gls{disnets}. The structure of the FCI is similar to that of the \gls{dci} signal, which is currently used in 5G NR to enable centralized scheduling~\cite{parkvall20185g}.
		\item We apply \gls{disnets} to the context of \gls{urllc} in \gls{iiot} environments. As such, we propose a new ad hoc traffic model in which industrial machines and users in a production line activate and generate traffic, respectively, based on some temporal and spatial correlations. This approach promotes more realistic, IIoT-specific simulations.
		\item We validate \gls{disnets} through \gls{e2e} simulations in terms of latency and reliability, against 5G NR \gls{gbs} and \gls{sps} baselines for resource allocation, and \gls{gfs} based on random access. Simulation results are given as a function of the number of \glspl{ue} in the network, the traffic configuration, and some other IIoT-specific system parameters. We show that \gls{disnets} achieves faster and more accurate resource allocation than its competitors, also in the presence of aperiodic and unpredictable traffic.
		
	\end{itemize}
	
	
	\subsection{Paper Organization}
	The rest of the paper is organized as follows. After discussing some related work in Sec.~\ref{sec:related_work}, in Sec.~\ref{sec:system_model} we describe our system model, in Sec.~\ref{sec:problem_formulation} we present our \gls{maccmab} problem formulation, in Sec.~\ref{sec:proposed_framework} we solve the problem using the proposed \gls{disnets} algorithm, in Sec.~\ref{sec:results} we provide numerical results, and in Sec.~\ref{sec:conclusion} we summarize our main conclusions and suggestions for future research.
	
	\section{Related Work}
	\label{sec:related_work}
	
	Achieving \gls{urllc} has been a long-standing research problem, motivated by the several services, applications, and verticals that pose stringent networking requirements~\cite{bennis:2018:urllc}. Adaptive modulation and coding schemes, short-length packet coding, channel access and re-transmission, and resource allocation are just a few examples of network operations that need to be optimized to achieve \gls{urllc}~\cite{net_latency_2019}. 
	Notably, as highlighted in~\cite{hong:2019:icc,globecom_mab,cuozzo2022enabling}, resource allocation currently represents the bottleneck to reduce the latency below $1$~ms. 
	This is mainly due to the current limitations of standard resource allocation protocols, which cannot dynamically and proactively trade off low latency and high reliability~\cite{parkvall20185g}, especially for uplink communication~\cite{cuozzo2022enabling}. 
	On one hand, centralized \gls{gbs} and \gls{sps} can coordinate the allocation of physical resources to minimize the number of collisions, but imply some sort of communication between the \glspl{ue} and the \gls{gnb} to agree on the resources to use before sending data~\cite{cuozzo2022enabling}. 
	On the other hand, distributed \gls{gfs} protocols, in which the \glspl{ue} can autonomously choose the physical resources for uplink communication without grants, can eliminate a good part of the delay. However, this option naturally leads to higher collisions probability as many \glspl{ue} may utilize the same wireless resources given the lack of coordination, which may prevent \gls{urllc} in dense networks~\cite{liu2020analyzing}. 
	
	
	Along these lines, the literature has proposed many solutions to optimize the system performance beyond model-based architectures, especially using \gls{ml} models trained on network data~\cite{she:2020:network_dl_survey, azari:2019:risk_urllc}. The general idea is to formulate the problem as a decision-making process, and to carefully design a feedback signal rewarding the agent as some network metrics (e.g., latency, reliability, fairness, power consumption, and so on) are satisfied. 
	For example, the authors in~\cite{kasgari:2019:icc_model_free} expressed the problem of resource allocation for \gls{urllc} as a \gls{mdp}, and solved it thorough \gls{drl}. 
	In \cite{gu:2021:knowledge_drl}, the authors introduced prior knowledge into the system, and showed that this approach can reduce the convergence time of the \gls{drl} solution. However, these works focus on downlink traffic with centralized scheduling, which is usually not scalable as the number of \glspl{ue} increases. 
	
	To target this problem, the research community has studied distributed multi-agent \gls{mdp}~\cite{bragato2023towards}, which allows the \glspl{ue}, i.e., the agents, to make autonomous decisions on the physical resource(s) to use for communication without the support from an external centralized entity~\cite{globecom_mab}.
	For instance, in \cite{bambos:2021:ra_bandit} the authors used game theory to define and study the performance of distributed resource allocation in which each agent is trained to select physical resources so as to avoid collisions. However, the problem was considerably simplified to be mathematically tractable, and no simulations on real communication systems have been performed. 
	In our recent paper~\cite{globecom_mab} we also proposed a similar formulation, in which four state-of-the-art \gls{mab} algorithms have been compared to identify the best implementation in terms of latency and reliability. The analysis suggested that the \gls{ts} algorithm is a good candidate, achieving zero collisions in our experiments.
	However, the system model in~\cite{globecom_mab} was quite simple, and most importantly each agent was designed to select only one orthogonal resource to transmit data, which is not realistic in practice. 
	In \cite{li:2018:vehicular_drl}, the authors developed a similar framework to manage distributed \gls{v2x} communication without relying on global information. In this case, each \gls{ue} was equipped with a \gls{drl} agent, which is more complex and difficult to handle than \gls{mab} but potentially offers more flexibility and better performance. However, each agent could still choose only one physical resource per scheduling opportunity, as choosing more would have increased exponentially the action space, so the complexity. 
	
	Based on the above introduction, in this work we decided to formulate the problem of distributed resource allocation as an \gls{maccmab} problem~\cite{intro_mab}, which generally achieves faster convergence than multi-agent \gls{drl}. 
	The \gls{maccmab} problem is solved using \gls{disnets}, which combines and extends the \gls{lts}~\cite{tutorial_thompson_sampling} and \gls{nlts}~\cite{Riquelme2018} algorithms so that agents can use (and are trained to choose) multiple orthogonal resources in the same scheduling unit, in the hope to transmit data faster. 
	We will demonstrate in Sec.~\ref{sec:results} that this approach outperforms state-of-the-art centralized and decentralized benchmarks.


	\section{System Model}
	\label{sec:system_model}
	
	In this section we present our system model. Specifically, we describe our factory layout and scenario in Sec.~\ref{sub:scenario}, the channel model in Sec.~\ref{sub:channel_model}, the traffic model to characterize IIoT-specific interactions between machines, end users, and the underlying factory geometry and functionalities in Sec.~\ref{subsec:traffic_model}, and the \gls{e2e} latency and reliability models in Sec.~\ref{subsec:latency}.
	
	\subsection{Scenario}
	\label{sub:scenario}
	
	\paragraph{Factory floor} 
	\label{par:factory_floor}

	\begin{figure}[t!]
		\begin{center}
			\includegraphics[width=0.99\columnwidth]{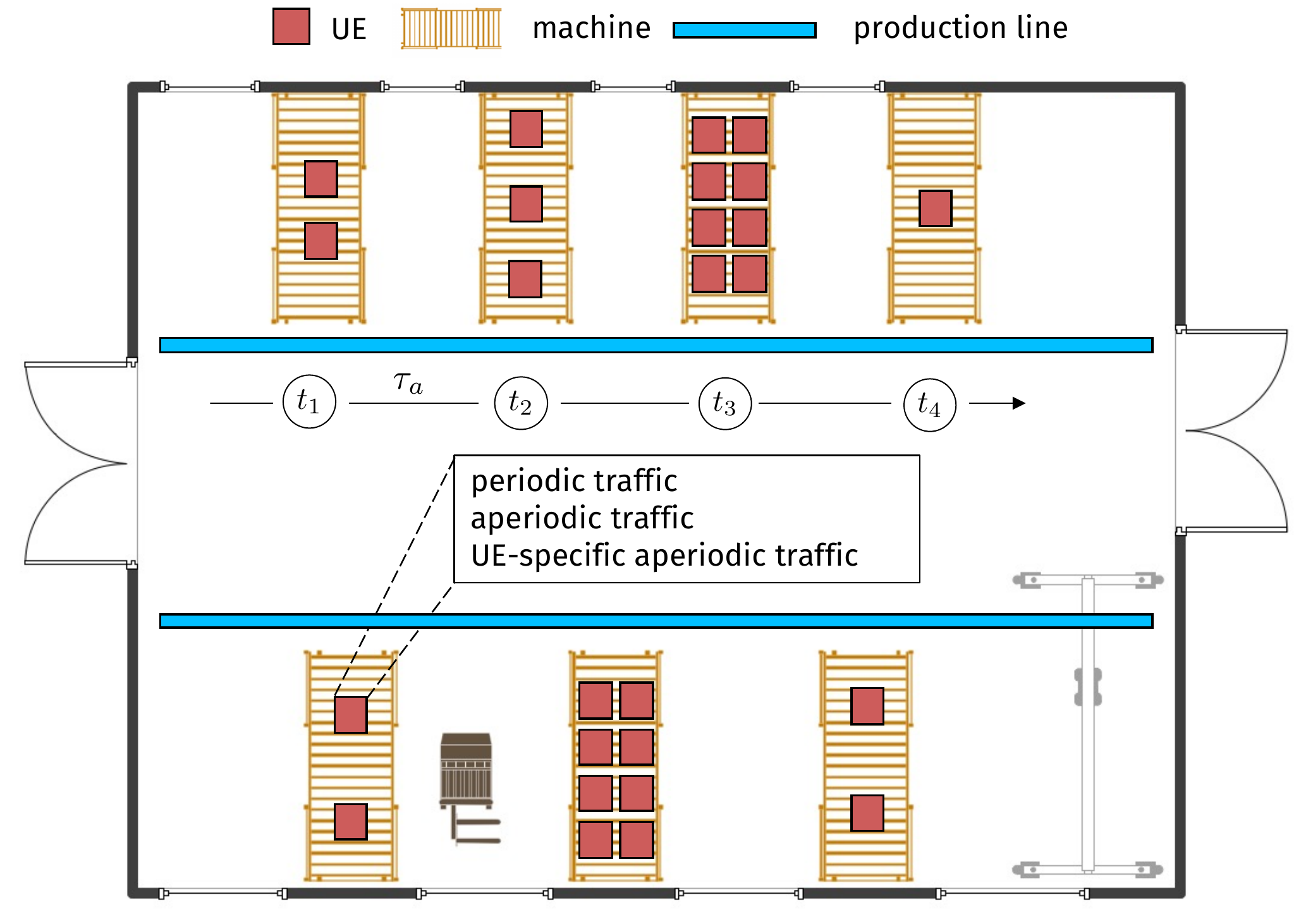}
			\caption{Factory floor layout (with $W=2$, $M=7$, and $N=18$) and traffic correlation. Specifically, machines in each production line are correlated, and activate according to a specific sequence on the production line, i.e., toward the right or the left. At $t_1$, $W=2$ machines (i.e., one per production line) activate, and the corresponding UEs onboard the active machines start sending data as periodic, aperiodic, or UE-specific aperiodic traffic. At $t_2=t_1+\tau_a$, these machines shut down and the next activation begins.\vspace{-0.55cm}}
			\label{fig:intermachine_correlation}
		\end{center}
	\end{figure}
	
	We consider a limited geographical area within an indoor factory floor, modeled as a parallelepiped of length $l$, width $w$, and height $h$, as reported in~\cite{5gaciarequirement}.
	Then, $M$ industrial machines are grouped into $W$ different production lines (each of which models an underlying industrial process), and are connected to a \gls{snpn}, i.e., a 5G remote and private network with a reserved \gls{ran} and \gls{5gc}~\cite{5gaciafactory}. 
	For example, Fig.~\ref{fig:intermachine_correlation} illustrates an example with $W=2$ production lines and $M=7$ machines.
	Machines are modeled as cubes of size $S$, and deployed across the factory floor according to a uniform distribution, ensuring a given inter-machine distance $D$ between the centers of the machines. 
	Onboard the machines, $N$ \glspl{ue} are distributed following the same method, at a maximum height $S$, and generate traffic according to pre-defined patterns (see Sec.~\ref{subsec:traffic_model}). 
	Moreover, obstacles act as obstructions between the UEs and the gNB.

	\paragraph{Resource allocation} 
	\label{par:resource_allocation}
	
	We consider uplink communication, i.e., from the $N$ \glspl{ue} to the \gls{cm}, that is a remote entity monitoring and controlling the machines from the \gls{5gc} through a \gls{gnb}.
	The total available bandwidth $B$ is split into $K$ orthogonal channels where, according to the 3GPP nomenclature, an orthogonal channel consists of $12$ \gls{ofdm} subcarriers. Time is also discretized into \glspl{su} whose duration is of $7$ \gls{ofdm} symbols. 
	Then, a \gls{rb} is defined as the minimum physical resource unit that can be allocated for data transmission, and consists of one orthogonal channel in frequency, and one \gls{su} in time. Within one \gls{su}, the first $4$ \gls{ofdm} symbols are dedicated to the \gls{pusch} used by the \glspl{ue} to transmit data, and the last $2$ \gls{ofdm} symbols are used by the \gls{gnb} to convey the \gls{fci}, as described in Sec.~\ref{sub:fci_context}. 
	
	\paragraph{Data transmission} 
	Whenever a \gls{ue} has new data packets to send, it can use multiple \glspl{rb} choosing different orthogonal channels within the same \gls{su}. 
	The assumption in our system model is that, whenever two or more \glspl{ue} use the same orthogonal channels in the same \gls{su}, i.e., the same set of \glspl{rb}, they create a collision, and we assume all data packets to be lost, i.e., they cannot be detected by the \gls{gnb}. 
	
	\subsection{Channel Model}
	\label{sub:channel_model}
	
	The channel is characterized based on the \gls{inf} model for \gls{iiot} networks~\cite[Table 7.2-4]{3gpp.38.901}. Specifically, the \gls{3gpp} identifies several \gls{inf} scenarios depending on the density of obstacles and the location of the \glspl{ue} with respect to the \gls{gnb}. Among these, in this paper we selected the most representative 3GPP InF scenario based on the 5G-ACIA factory layout and geometry described in~\cite{5gaciafactory} and Sec.~\ref{sub:scenario}. 
	
	The path loss depends on the \gls{los} or \gls{nlos} condition of the channel. In this regard, the 3GPP provides an expression for the \gls{los} probability in~\cite{3gpp.38.901}. However, we do not adopt a statistical model to discriminate between \gls{los} and \gls{nlos} propagation. On the contrary, we implement a geometry-based approach that checks whether the joining line between the UE's and the gNB's centers intersects one or more obstacles. If so, the UE is considered in NLOS, otherwise it is in LOS. 
	Then, the quality of the received signal is assessed in terms of the \gls{snr}, which is defined as
	\begin{equation}
		\label{eq:snr_definition}
		\text{SNR} = \frac{P_{\rm TX} \cdot G_{\rm UE} \cdot G_{\rm gNB}}{PL \cdot P_N},
	\end{equation}
	where $P_{\rm TX}$ is the transmit power, $G_{\rm UE}$ and $G_{\rm gNB}$ are the antenna gains at the UE and the gNB, respectively, $PL$ is the path loss, and $P_N$ is the \gls{awgn} power. The latter is computed as $k_B \cdot T \cdot B$, where $k_B$ is the Boltzmann constant, $T_B$ is the system noise temperature (in K), and $B$ is the total bandwidth (in Hz) at the \acrshort{gnb}.
	The \gls{snr} is used to check whether a data block is correctly decoded, i.e., if the \gls{snr} is above a given threshold SNR$_{\rm th}$, and to determine the modulation order to be used for data transmission according to~\cite{38214}.

	\subsection{The Spatio-Temporal Correlated Traffic Model}
	\label{subsec:traffic_model}
	
	In most of the literature, uplink traffic is assumed either periodic (i.e., UEs generate data at predefined time intervals) or totally aperiodic (i.e., UEs generate data at random intervals), and machines in the factory floor can generate packets simultaneously. In this paper, we propose an alternative traffic model where data packets are generated according to pre-defined 
	statistics, to better characterize \gls{iiot} interactions.
	In this model, machines within a production line are sequentially activated, emulating the workflow of the corresponding industrial process. Then, \glspl{ue} onboard active machines produce data traffic (either periodic or aperiodic~\cite{5gaciaarchitecture}) for an entire ``activation period'' of duration $\tau_{a}$, after which another machine in the current production line will activate.
	
	As such, the traffic model accounts for both spatial and temporal correlation, as illustrated in Fig.~\ref{fig:intermachine_correlation}. Notably, it is possible to identify two different types of correlation.

	\paragraph{Inter-machine correlation} It refers to the way machines activate. Specifically, when an event occurs, one machine per production line is activated. Then, after  the activation period, the next machine activates following the flow of the production line, at a sequence that depends on the factory geometry and the number of machines and lines.
	
	\paragraph{Intra-machine correlation} It refers to the way \glspl{ue} onboard the active machines generate packets. After one machine per line is activated, the \glspl{ue} associated with those machines activate too. Then, active \glspl{ue} generate a flow of packets according to some statistics, e.g., in terms of the inter-packet interval and/or the packet size, until some new machines in the production line activate. We consider:
	\begin{itemize}
		\item \emph{Periodic traffic}, if packets are generated at constant periodicity $\tau$~\cite{5gaciaarchitecture}. For example, UEs periodically measure and report physical parameters (e.g., temperature, pressure, radiation) from the production process.
		\item \emph{Uniformly aperiodic traffic}, if the inter-packet interval $\tau$ is modeled as a uniform random variable in $[t_{min}, t_{max}]$. For example, UEs make unscheduled aperiodic transmissions in case abnormal measurements are detected.
		\item \emph{UE-specific aperiodic traffic}, in which the extreme values $t_{min}$ and $t_{max}$ are \gls{ue}-dependent parameters. Specifically, we now assume that \gls{ue}$_n$, $\forall n \in \mathcal{N}$, where $\mathcal{N}$ is the set of \glspl{ue}, generates with probability one another packet in the interval $[t^n_{min}, t^n_{max}]$, with $t_{min}^n$ and $t_{max}^n$ modeled as uniform random variables within the intervals $[t_{min}, t_{max}]$ and $[t_{min}^n, t_{max}]$, respectively. Notably, we now have $t_{min} \leq t^n_{min} \leq t^n_{max} \leq t_{max}$. The rationale behind this new traffic model is that, in real-world factories, some sensors/UEs may control different parts or mechanisms of the same inter-machine process, thus activating with statistics that depend on their roles or position, and so are UE-specific.  
	\end{itemize}
	
	\subsection{Latency and Reliability Models}
	\label{subsec:latency}
	
	\paragraph{Latency}
	We require our system to minimize the \gls{e2e} latency in uplink, which is defined as the time from when one packet is generated at the UE's application to when the same packet is successfully received by the \gls{cm}. Specifically, the \gls{e2e} latency $L$ of a packet is computed as:
	\begin{equation}
		\label{eq:e2e_latency}
		L = T_P + T_{\rm RAN} + T_{\rm TX} + \tau_P + T_{\rm DAS} + \tau_F + T_{\rm gNB} + T_{\rm CN},
	\end{equation}
	where, based on the analysis in~\cite{cuozzo2022enabling}:
	
	\begin{itemize}
		\item $T_P$ is the time for the \gls{ue} to create the data packet, i.e., to add headers across the 5G protocol stack;
		\item $T_{\rm RAN}$ is the time between the generation of the data packet at the \gls{phy} layer and the packet transmission, which depends on the scheduling algorithm;
		\item $T_{\rm TX}$ is the transmission time;
		\item $\tau_P$ is the propagation time from the \gls{ue} to the fronthaul of the \acrshort{gnb}, i.e., the \gls{das};
		\item $T_{\rm DAS}$ is the time for the \gls{das} to process the received data packet, and to send it to the \gls{gnb};
		\item $\tau_F$ is the time for the signal to travel from the \gls{das} to the \acrshort{gnb}, generally through a high-capacity optical fiber;
		\item $T_{\rm gNB}$ is the time for the \acrshort{gnb} to process the received data block, and to send it to the \gls{cm};
		\item $T_{\rm CN}$ is the delay introduced by the \gls{5gc}, that is the time for the message to reach the \acrshort{cm} from the \acrshort{gnb}.
	\end{itemize}
	
	Finally, we denote as $\bar{L}$ the average \gls{e2e} latency, averaged over the data packets generated by the \glspl{ue} within the simulation time $T_S$, and over the number of \glspl{ue}.
	
	\paragraph{Reliability}
	We also require our system to operate with high reliability. We define 
	a reliability metric $\eta_t(L_{\text{th}})$ as the empirical probability that the \gls{e2e} latency of a packet is below a pre-defined threshold $L_{\text{th}}$ during \gls{su}$_t$, and $\bar{\eta}_t(L_{\text{th}})$ is the empirical average of $\eta_t(L_{\text{th}})$ within the simulation time~$T_S$. 

	\section{Problem Formulation}
	\label{sec:problem_formulation}
	
	The aim of our work is to minimize $T_{\rm RAN}$ in Eq. \eqref{eq:e2e_latency}, which dominates the \gls{e2e} latency, and depends on the underlying resource allocation procedure.
	As mentioned in Sec.~\ref{sec:introduction}, standard 5G NR protocols mainly adopt either centralized scheduling at the \gls{gnb} (i.e., \gls{gbs} and \gls{sps}), which introduces delays due to the rigidity of the  resource allocation scheme with respect to the traffic generation process, or distributed scheduling (i.e., \gls{gfs}), which may result in collisions.
	In turn, we propose \gls{disnets}, a new scheduling framework that combines the benefits of the two: on one side, resource allocation is decentralized, in the sense that \glspl{ue} autonomously decide how to allocate resources without significant interactions with the \gls{gnb}, which eliminates the waiting time to receive scheduling grants; at the same time, \glspl{ue} exploit \gls{ml} to optimze scheduling decisions based on traffic correlations, which may reduce the probability of collision. Our research problem is formulated as a \gls{maccmab} problem, as described below.
	
	\subsection{The \acrshort{ccmab} Problem}
	\label{sub:ccmab}
	
	The problem formulation is built on top of the \gls{ccmab} framework~\cite{NEURIPS2018_207f8801}. Specifically, every time UE$_n\in \mathcal{N}$, i.e., an agent, has data to send in \gls{su}$_t$, it will autonomously choose the physical resources to be used for transmission. 
	The total available bandwidth is split into $K$ orthogonal channels, and we denote with $\mathcal{K} = \{1, 2, \ldots, K \}$ the set of channels. In \gls{ccmab} parlance, the $K$ orthogonal channels are the $K$ feasible actions that can be chosen by the agent. 
	To take an action, each agent can rely on side information (i.e., the context $s_t \in \mathcal{S}$) that describes the state of the environment (i.e., the wireless network) in \gls{su}$_t$, which we model as a random variable sampled according to the system's probability $P_S$. Given the context $s_t$ and the action $k_t$ chosen by the agent in SU$_t$, the environment returns a reward $r_t \in [-1, 1]$ according to the probability $P_R(s_t, k_t)$, which reflects the probability that data transmission using channel $k_t$ in context $s_t$ was successful.
	Specifically, $r_t=-1$ if collisions happen, otherwise it is proportional to the number of transmitted bits (see Eq.~\eqref{eq:rewards-dcnlts} for further details). 
	Then, $\mu(s, k)$ is the average reward 
	with respect to the distribution $P_R(s, k) \; \forall s \in \mathcal{S}, \forall k \in \mathcal{K}$. 
	
	In our framework, the \gls{ccmab} problem is extended by allowing agents to take more than one action in each \gls{su}, i.e., using multiple orthogonal channels in parallel in the same \gls{su}, which is important to provide \gls{urllc}. 
	Therefore, we define a super-action $\theta_t \subset \mathcal{K}$ as a set of actions in SU$_t$,  so $\theta_t$ is an element of the super-set $\Theta$ of $\mathcal{K}$, i.e., the set of all possible subsets of $\mathcal{K}$. 
	The reward $r_t$ is then sampled according to $P_R(s_t, \theta_t)$. 
	Interestingly, we can exploit the structure of the environment to assume that 
	\begin{align}
		\mu(s, \theta) \sim \sum_{k \in \theta} \mu(s, k), \quad \forall s \in \mathcal{S}, \;  \theta \subset \mathcal{K},
		\label{eq:super-reward}
	\end{align} 
	i.e., the average reward relative to super-action $\theta$ is proportional to the sum of the average rewards of the single actions. 

	To choose the super-action in SU$_t$, the agent employs a policy ${\pi_t: \mathcal{H}^{t-1} \times \mathcal{S} \rightarrow \Phi(\Theta)}$, which is a map from the history $H_t = \{ (s_1, \theta_1, r_1), \ldots, (s_{t-1}, \theta_{t-1}, r_{t-1})\} \in \mathcal{H}^{t-1}$ of previous contexts, actions, and rewards, to a probability distribution over the set of feasible super-actions $\Theta$. 
	Given a horizon $T$, the goal of the agent is to find the policy $\pi^*$ that maximizes the expected sum of rewards over time,~i.e.,
	\begin{align}
		\label{eq:opt_policy}
		\pi^* = \argmax_{\pi} \mathbb{E}\left[ \sum_{t=1}^{T} \mu(s_t, \theta_t)\right],
	\end{align}
	where the expectation is taken with respect to the distributions of $P_R$ and $P_S$ and the agent's policy $\pi_t$, used to sample super-actions according to $\theta_t \sim \pi_t(s_t)$. 
	
	\subsection{The \gls{maccmab} Problem}
	\label{sub:ma_ccmab}

	\begin{figure*}
		\centering
		\includegraphics[width=0.86\textwidth]{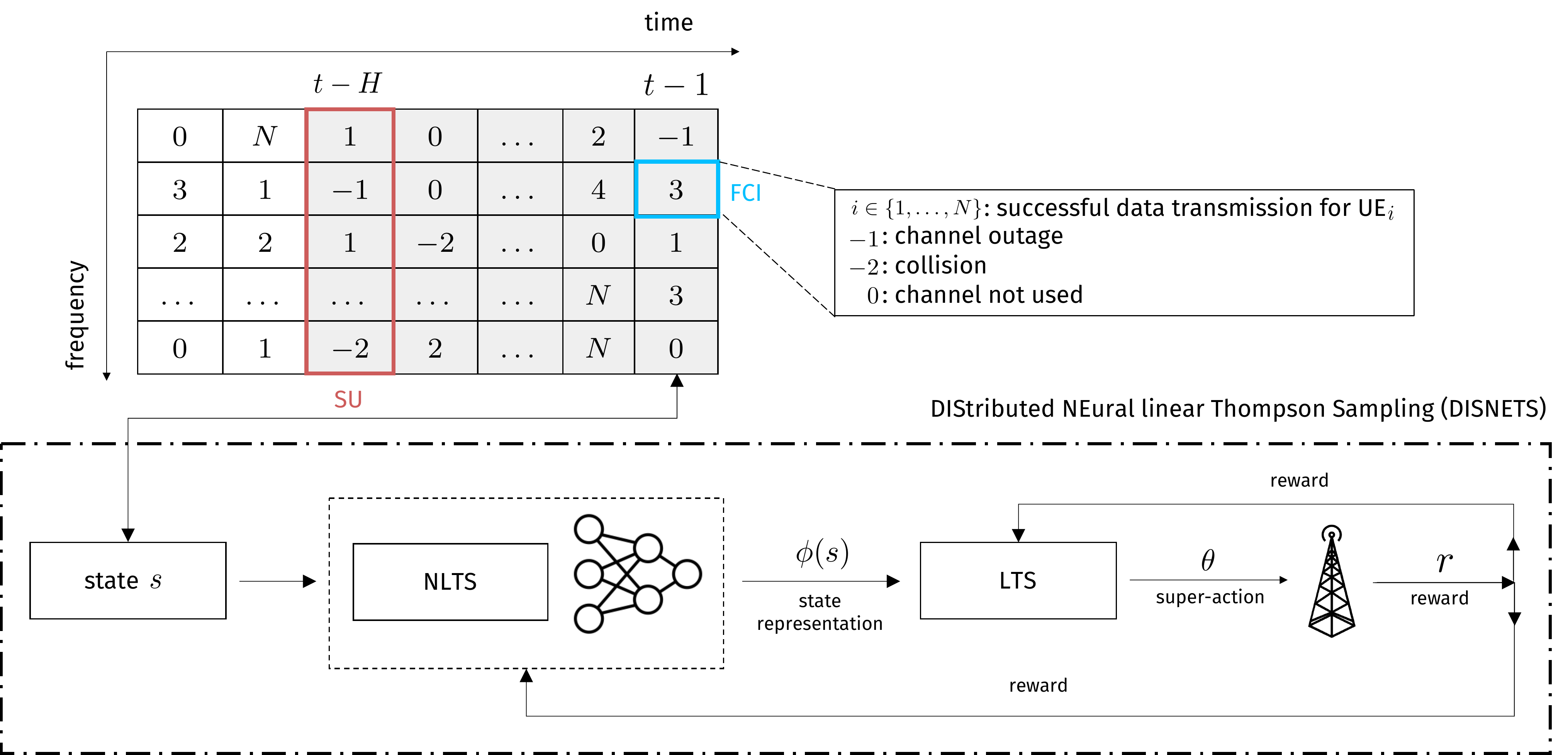}
		\caption{Schematic representation of \gls{disnets}. The framework consists of (i) the state/context $s$, (ii) an \gls{nlts} module to provide the non-linear representation of the context $\phi_{\omega}(s)$, (iii) an \gls{lts} module to choose a super-action $\theta\in\mathcal{K}$ corresponding to the set of orthogonal channels to use to transmit data, (iv) the reward $r$ (incorporated within the FCI) to update the \gls{nlts} and \gls{lts} parameters.\vspace{-0.5cm}}
		\label{fig:disnets}
	\end{figure*}

	In our work we further extend the \gls{ccmab} problem in Sec.~\ref{sub:ccmab}, and formulate a new \gls{maccmab} problem where the reward depends on the other agents' actions in \gls{su}$_t$. Therefore, the reward $r_{n,t}$, the context $s_{n,t} \in \mathcal{S}_n$, and the policy $\pi_{n, t}$ at time $t$ are now a function of the agent $n$.
	This is critical to support more accurate distributed resource allocation, where there is no or little interaction among the~\glspl{ue}.
	
	A collision event is expressed by the random variable $\chi(k, t) \in \{0,1\}$, which is equal to $1$ when collision(s) happen in the orthogonal channel $k$ in \gls{su}$_t$, and $0$ otherwise. As such, it is modeled by each agent as a Bernoulli random variable of parameter $\varphi(s_{n,t})$.\footnote{Notice that $\chi(k, t)$ does not explicitly depend on $n$ but has the same value for all the agents. In turn, its statistics, specifically $\varphi(s_{n,t})$, are modeled using the local context $s_{n, t}$ as agents have missing information on the other agents'~actions.}  
	By modeling assumption, whenever two agents $n$ and $n'$ play super-actions $\theta_{n,t}$ and $\theta_{n',t}$, respectively, we have that $\chi(k, t) = 1, \; \forall s \in \mathcal{S}, \; \text{if } k \in \theta_{n,t} \cap \theta_{n',t}$, as collisions occur.
	Notice that agents make decisions based only on local data, so that agent $n$ does not know a priori the structure of vectors $\theta_{n',t}$, $n'\neq n$. As such, agents can just rely on the local context $s_{n, t}$, and the statistical knowledge  acquired through history $H_{n, t}$.
	
	\section{Proposed Solution: the \gls{disnets} Framework}
	\label{sec:proposed_framework}
	
	In this section, we present the proposed solution to the \gls{maccmab} problem introduced in Sec.~\ref{sub:ma_ccmab}.
	The proposed framework implements (i) a new control signal called \gls{fci} (Sec.~\ref{sub:fci_context}), which conveys the information to compute the reward $r_{n, t}$ and is used to generate the context $s_{n,t}$, and (ii) the \gls{disnets} algorithm (Sec.~\ref{sub:nlts}), which combines \gls{lts} and \gls{nlts} to solve Eq.~(\ref{eq:opt_policy}), and allows end users to autonomously allocate resources for uplink transmissions.
	
	\subsection{Context and \gls{fci} Implementation}
	\label{sub:fci_context}
	
	As introduced in Sec.~\ref{sub:ccmab}, the agents optimize  policies $\pi_{n, t}$ based on the rewards $r_{n,t}$, $\forall n\in \mathcal{N}$.
	In our framework, the reward is based on the new \gls{fci} signal depicted in Fig.~\ref{fig:disnets}, which has a similar structure as the 5G NR \gls{dci} signal.
	The \gls{fci}, sent from the gNB to the UEs, is located in the last $2$ \gls{ofdm} symbols of each \gls{su}, and includes the transmission outcomes in that \gls{su} relative to each orthogonal channel. 
	Based on the received power on each orthogonal channel, the gNB can distinguish
	among four outcomes, namely successful transmission for UE$_i$ (FCI $=i$, $i\in\{1,\dots,N\}$), outage (FCI $=-1$), collision (FCI $=-2$), or channel not used (FCI $=0$). 
	
	Based on the \gls{fci}, the \glspl{ue} can compute their local contexts $s_{n,t}$, $\forall n\in \mathcal{N}$, which are used to statistically describe the state of the system in \gls{su}$_t$. Specifically, the context $s_{n,t}$ aggregates the outcomes from $H$ previous FCI signals, which provides a history of previous resource allocation decisions.
	

	\subsection{\gls{disnets} Implementation}
	\label{sub:nlts}
	
	In this section we describe our \gls{disnets} framework, represented in Fig.~\ref{fig:disnets}, to solve the \gls{maccmab} and support distributed resource allocation. \gls{disnets} extends the \gls{nlts} algorithm, which is in turn based on the \gls{lts} algorithm. 
	
	\emph{Notation.} For simplicity, we drop subscript $n$, with $n\in\{1,\dots,N\}$, to represent the index of the agent/UE.
	
	\paragraph{\acrfull{lts}}
	\label{subsub:linear_ts}
	\gls{disnets} is built on top of the \gls{lts} algorithm~\cite{Agrawal2013} to choose an action $k \in \mathcal{K}$, which returns the orthogonal channel that the UE can use to transmit data. 
	\Gls{lts} assumes that the average reward behind each action $k \in \mathcal{K}$ is a linear function of the context $s_t \in \mathcal{S}$, and of an unknown action parameter vector $\beta_k$, i.e., $\mu(s, k) = \langle s, \beta_k \rangle $. 
	In this case, estimating the most accurate vector $\beta_k$, $\forall k \in \mathcal{K}$, turns out to be an {online} linear regression problem. 
	The problem is online because the target values, i.e., the rewards associated to the pair $(s, \beta_k)$, are observed by the agent as it interacts with the environment and takes actions, respectively, and are not available at the beginning of the training process like in supervised learning.

	To solve this online problem, \gls{lts} assumes that the rewards, given the context $s$ and action $k$, are modeled as a Gaussian random variable\footnote{According to~\cite{Agrawal2013}, the rewards are not required to be Gaussian to converge to the optimal actions, but their domains need to be bounded.} $R(s, k)$,  i.e.,  $R(s, k) \sim \mathcal{N} \left( s^T \beta_k, \nu_k^2\right)$. 
	The \gls{lts} algorithm maintains a distribution of the parameter vector of each action $k$ at time $t$, i.e., $\beta_k(t)$. Therefore $P(\beta_k(t)) \propto \mathcal{N} \left( \hat{\beta}_k(t), \nu_k^2 \left( \Phi_k(0) + \Phi_k(t)\right)^{-1} \right)$, where 
	\begin{align}
		\Phi_k(0) & = \lambda \cdot I_d; \\
		\Phi_k(t) &= \sum_{\tau = 1}^{t-1} s_\tau s_\tau^T \cdot \mathbf{1}_{k_t = k};\\
		\hat{\beta}_k(t) & = (\Phi_k(0) + \Phi_k(t))^{-1} \sum_{\tau=1}^{t-1} s_\tau r_\tau \cdot \mathbf{1}_{k_t = a}.
	\end{align}
	In particular, $\lambda$ is a hyper-parameter governing the initial exploration, $I_d$ is the identity matrix of size $d$, and $\mathbf{1}_{k_t = k}$ is the indicator function and is equal to 1 if $k_t = k$ or 0 otherwise. After observing context $s_t$ and taking action $k_t$, the agent receives a reward $r_t$ based on the FCI (Sec.~\ref{sub:fci_context}), and updates the distribution of the parameter vector at $t+1$~as
	\begin{align}
		\begin{split}
			\label{eq:posterior}
			& P(\beta_k(t+1) | s_t, r_t) \\
			&\propto P(r_t | s_t, \beta_k(t+1)) \cdot P(\beta_k(t+1)) \\
			& \propto \mathcal{N} \left( \hat{\beta}_k(t+1), \nu_k^2 (\Phi_k(0) + \Phi_k(t+1))^{-1} \right).
		\end{split}
	\end{align}
	Based on the {posterior update rule} in Eq.~\eqref{eq:posterior},
	the agent samples $K$ vectors $\hat{\beta}_k$, $\forall k \in \mathcal{K}$, and plays action $k_t = \argmax_{k \in \mathcal{K}}\langle s_t, \hat{\beta}_k\rangle$. 
	The interesting property of \gls{lts} is that it balances
	\emph{exploration}, i.e., sampling random actions to explore their reward statistics, and \emph{exploitation}, i.e., exploiting the knowledge collected in previous time slots to optimize future decisions, and presents good empirical performance~\cite{globecom_mab}. 
	
	\begin{mytheo}{\gls{lts}}{theoexample}
		\emph{Objective.} Choose an action $k\in\mathcal{K}$, and return the orthogonal channel that a UE should use to transmit~data. 
		
		\emph{Problem.} It works under the assumption of linear dependency between the context and the average reward obtained by playing action $k\in\mathcal{K}$.
	\end{mytheo}
	
	\paragraph{\acrfull{nlts}}
	\label{subsub:neural_linear_ts}
	
	
	As discussed in the previous paragraph, \gls{lts} assumes a linear relation between the context and the average reward obtained by playing one specific action. However, linear relations are not complex enough to model real-world scenarios.
	This problem is usually solved using \gls{drl}, which requires a long training phase. 
	On the other hand, \glspl{dnn} are potentially good candidates to model the many different relations between contexts and actions. 
	On the downside, when using \glspl{dnn} there are no closed form solutions for the posterior updates, like the one in Eq.~(\ref{eq:posterior}). As such, different approximate solutions have been proposed in the literature. 
	
	In particular, the authors in~\cite{Riquelme2018} introduced a new algorithm, called \gls{nlts}, that models the non-linearity between contexts and rewards by assuming that the average reward $\mu(s, k)$ is equal to a linear combination of the action parameter vector $\beta_k$ and a non-linear representation of the context $\phi(s)$, i.e.,  $\mu(s, k) = \langle \phi(s), \beta_k\rangle$. 
	To do so, a \gls{dnn} $f_{\omega}(s)$: $\mathcal{S} \rightarrow \mathcal{R}^K$, parameterized by the weights vector $\omega \in \Omega$, is trained to estimate the average reward for each action, where $\mathcal{R}$ is the range of the actions' reward, assumed to be the same for all actions. 
	The non-linear representation of the context $\phi(s)$, or $\phi_{\omega}(s)$ to illustrate the dependency on $\omega$, is the output of the last hidden layer of the \gls{dnn}, i.e., the input of the last layer, whose output is $f_{\omega}(s)$.
	Then, NLTS uses \gls{lts}  to choose the orthogonal channel for the \glspl{ue} to transmit data, working on the state representation $\phi_\omega(s)$ in place of $s$.
	
	We highlight that the posteriors of \gls{lts} are updated based on Eq.~\eqref{eq:posterior} every time the agent makes a decision, while the parameters $\omega$ of the \gls{dnn} are updated at fixed intervals of $O$ steps.\footnote{It is important to properly balance the optimization steps of the \gls{lts} module (performed at each interaction with the environment) and that of the \gls{nlts} module (performed every $O$ interactions with the environment) of \gls{disnets} to avoid training instabilities, as one module depends on the~other.} 
	Another feature introduced by \gls{nlts} is that the distribution of the noise variance of the reward $\nu_k$ is modeled using the {Inverse Gamma} distribution, i.e., $\nu_k(t) \sim\text{IG}(a_k(t), b_k(t))$. 
	The pseudocode of \gls{nlts} is reported in Algorithm~\ref{alg:neural_linear_ts}.
	
	\begin{algorithm}[t!]
		\textbf{Initialize} $\Phi_k(0) = \lambda \cdot I_d$, $\hat{\beta}_k(0)=\beta_k(0) = 0$, $\psi_k = 0$ \\
		\ForEach{$t \in 1, \dots, T$}
		{
			Observe $s_t$ and compute $z_t = \phi_{\omega}(s_t)$\\
			Sample w.r.t $\nu_k(t)$, $\forall k \in \mathcal{K}$, from $\text{IG}(a_k(t), b_k(t))$ \\
			Sample w.r.t $\beta_k$, $\forall k \in \mathcal{K}$, from $\mathcal{N}(\hat{\beta}_k(t), \nu_k^2 \left( \Phi_k(0) + \Phi_k(t)\right)^{-1})$ \\
			Play $k_t = \argmax_{k \in \mathcal{K}} z_t^T \beta_k$\\
			Observe $r_t$ and store $\left(s_t, k_t, r_t\right)$ in the buffer \\
			Update action posterior (Eq.~(\ref{eq:posterior})), using context $z_t$\\
			Update noise posterior \\
			\If{$\mod(t,O) = 0$}{
				Train $f_{\omega}$ with \textbf{SGD} using samples in the buffer \\
				Compute new $z_t$, and update \gls{lts}\\
			}
		}
		\caption{Neural Linear \acrlong{ts}}
		\label{alg:neural_linear_ts}
	\end{algorithm}

	\begin{mytheo}{\gls{nlts}}{theoexample}
		\emph{Objective.} Compute the non-linear representation of the context $\phi_{\omega}(s)$, i.e., the output of the last hidden layer of the \gls{dnn}, and use \gls{lts} to choose an action $k\in\mathcal{K}$, which returns the orthogonal channel that each UE should use to transmit data.
		
		\emph{Problem.} The agent can only play single actions, i.e., each UE can transmit through one orthogonal channel.
	\end{mytheo}
	
	\paragraph{\gls{disnets} (Proposed)}
	\label{subsub:combinatorial_ts}
	
	As mentioned, we now enhance the \gls{nlts} algorithm to allow each agent to choose multiple actions, i.e., to use multiple orthogonal channels in the same \gls{su} as expected in real systems, which defines the \gls{maccmab} problem in Sec.~\ref{sub:ma_ccmab}. 
	Intuitively, there are two main issues when applying basic \gls{nlts} to \gls{maccmab}: first, the super-action $\theta$ that maximizes the expected reward $\mu(s, \theta)$ is unknown a priori, as the effects of single actions can be combined in different ways to obtain the super-action's reward~\cite{Wang:icml:2018}; second, the complexity of combining super-actions increases exponentially with the number of single actions, which is not tractable.
	
	We design the reward for a single action $k \in \mathcal{K}$ in \gls{su}$_t$ as
	\begin{align}
		\label{eq:rewards-dcnlts}
		r_t & =  
		\begin{cases} 
			\bar{\rho}_{k}(t) & {\rm if }\; \chi(k, t) = 0; \\
			-1 & {\rm if }\; \chi(k, t) = 1,  
		\end{cases}
	\end{align}
	where $\bar{\rho}_{k}(t) \in [0, 1]$ represents the total number of bytes that can be sent during \gls{su}$_t$ using channel~$k$, normalized by the maximum number of bytes that can be sent when using the maximum modulation order, and $\chi(k, t) \in \{0,1\}$ indicates whether a collision happens using channel $k$ during SU$_t$. As such, the reward in Eq.~\eqref{eq:rewards-dcnlts} is a function of both the channel and the resource allocation policies of all the agents.
	Furthermore, based on the assumption in Eq.~\eqref{eq:super-reward}, we have that the average reward of super-action $\theta$ is the sum of the single average rewards, i.e., 
	\begin{equation}
		\label{eq:total_reward}
		\mu(s, \theta) 
		= \sum_{k \in \theta} \langle \phi_{\omega^*}(s), \beta_k \rangle, \:  \forall s \in \mathcal{S}, \: \forall \theta \in \Theta,
	\end{equation}
	where we assume that there exists a DNN $\omega^* \in \Omega$ that provides the exact non-linear representation of the context $s$, $ \forall s \in \mathcal{S}$. 
	The reward in Eq.~\eqref{eq:total_reward} leads to a particular case of \gls{maccmab} referred to as \emph{matroid bandits}~\cite{Kveton2014}. In this case, it is possible to estimate the average reward obtained by super-action $\theta$ as the sum of the estimated average rewards of all its base actions. 
	
	Based on the above introduction, we extend \gls{nlts} into the proposed \gls{disnets} algorithm so that the agent can play super-action $\theta_t$ in SU$_t$ based on the following criteria:
	\begin{enumerate}
		\item  Sample $K$ vectors $\{ \hat{\beta}_k \}_{k=1}^K$ as described in Sec.~\ref{sub:nlts};
		\item Compute the non-linear representation of the context, i.e., $\phi_{\omega}(s_t)$, based on the context $s_t$ at SU$_t$;
		\item Take super-action $\theta_t = \{ k \in \mathcal{K} \;  : \; \phi_{\omega}(s_t)^T \beta_k > \epsilon \}$, i.e., the agent transmits using all the orthogonal channels whose estimated reward (which is an indication of the transmission data rate) is larger than $\epsilon$. 
	\end{enumerate}

	Consequently, the average number of orthogonal channels that an agent can use is not constant, but rather learned and adjusted via \gls{disnets} given the reward history and the context at time $t$. 
	In addition, we further introduced a {variance decaying}~factor~$\gamma$ to scale the sampling variance of the reward $\nu_k(t)$ by $\gamma$, in order to force \gls{disnets} to become more deterministic as the training progresses.
	
	\begin{mytheo}{\gls{disnets}}{theoexample}
		\emph{Objective.} Compute the non-linear representation of the context $\phi_{\omega}(s)$, and choose a super-action $\theta_t\in\mathcal{K}$ which returns the set of orthogonal channels that each UE should use to transmit data.
	\end{mytheo}

	\section{Numerical Results}
	\label{sec:results}

	In Sec.~\ref{sub:alg_design} we present our simulation parameters, in Sec.~\ref{sub:num_traninig} we show the convergence performance of \gls{disnets}, and in Secs.~\ref{sec:overhead} and \ref{sub:performance_evaluation} we evaluate the performance of \gls{disnets} against some other benchmarks in terms of overhead, latency, and reliability. 
	
	\subsection{Simulation Parameters}
	\label{sub:alg_design}
	
	Simulation parameters are reported in Table~\ref{tab:system_parameter_settings}.

	\begin{table}[!t]
		\centering
		\renewcommand{\arraystretch}{1.2}
		\footnotesize
		\caption{Simulation parameters.}
		\label{tab:system_parameter_settings}
		\begin{tabular}{|l|l|}
			\hline
			{Parameter} & {Value}  \\\hline
			Carrier frequency ($f_c$)& 3.5 GHz\\ 
			Overall system bandwidth ($B$) & $60$ MHz \\
			\gls{5g} protocol stack header ($H$) & 72 bytes~\cite{etsi_tr_137_901_5}\\
			Subcarrier spacing ($\Delta f$) & $60$ kHz\\
			SNR threshold (SNR$_{\rm th}$) & $-5$ dB  \\   
			Latency threshold ($L_{\rm th}$) & $1$ ms  \\   
			Noise temperature ($T_B$) & 290 K   \\  
			Antenna gain ($G_{\rm UE}=G_{\rm gNB}$) & 0 dB \\   
			UE (UL) transmit power ($P_{\rm TX, UL}$) & 23 dBm \\   
			gNB (DL) transmit power ($P_{\rm TX, DL}$) & 30 dBm \\  \hline
			Processing time at the UE ($T_P$) & 7 OFDM symbols   \\  
			Processing time at the UE ($T_{\rm gNB}$) & 7 OFDM symbols   \\
			\gls{5gc} delay ($T_{\rm CN}$) & 0.1 ms \\    
			DAS delay ($T_{\rm DAS}$) & 0.05 ms \\  \hline
			Length of the factory floor ($l$) & $20$ m~\cite{5gaciaarchitecture} \\
			Width of the factory floor ($w$) & $20$ m\\  
			Height of the factory floor ($h$) & $4$ m \\  
			Inter-machine distance ($D$) & 5 m \\   
			Side of the machine ($S$) & 3 m \\  
			Number of production lines ($W$) & 4 \\   
			Number of machines ($M$) & 4/line \\   
			Inter-machine activation period $\tau_a$ & $8$ ms  \\
			Packet size ($Z_p$) & $616$ Bytes \\ 
			Simulation time ($T_S$) & 7 s \\\hline
		\end{tabular}
	\end{table}

	\paragraph{System parameters} 
	\label{par:parameters}
	The system operates with a carrier frequency of $f_c=3.5$ GHz and a bandwidth of $B=60$ MHz. We set 3GPP NR numerology 2 (i.e., a subcarrier spacing of $\Delta_f=60$ KHz), which leads to 84 \glspl{rb}~\cite{38211}.
	For the latency in Eq.~\eqref{eq:e2e_latency}, according to the 5G standard specifications we assume that (i) the processing times $T_P$ and $T_{\rm gNB}$ at the \glspl{ue} and the \gls{gnb}, respectively, are both equal to $7$ \gls{ofdm} symbols, (i.e., $116.9$ $\mu$s for numerology~2), (ii) the propagation time $\tau_{P}$ is neglected because it can be compensated with an accurate timing advance technique (see \cite{3gpp.38.825}), and (iii) $\tau_{F}$ is also neglected due to its minor impact on~$L$. 
	
	\paragraph{\gls{disnets} parameters} 
	\label{par:disnets_parameters}
	
	\begin{table}[t!]
		\footnotesize
		\renewcommand{\arraystretch}{1.4}
		\caption{Structure of the DNN used in the \gls{disnets} algorithm. }
		\label{tab:network_structure}
		\centering
		\begin{tabular}{|l|l|l|l|l|}
			\hline
			\centering
			& {Type} & {Size}  & {Max Pool}  & {Activation} \\
			\hline
			{Layer 1} & Conv. & (10, 4, 4) & (3, 3)  & Leaky ReLu \\
			{Layer 2} & Conv. & (10, 3, 3) & (3, 3)  & Leaky ReLu \\
			{Latent Layer} & Linear & $10$ &  $\cdot$ & Leaky ReLu \\
			{Output Layer} & Linear & $K$ & $\cdot$  & Identity \\\hline
			\multicolumn{3}{|c|}{Variance decaying factor ($\gamma$)} & \multicolumn{2}{c|}{$0.9999$}\\
			\multicolumn{3}{|c|}{Number of optimization steps ($O$)} & \multicolumn{2}{c|}{$100$}\\
			\hline
		\end{tabular}
	\end{table}
	
	The configuration of the \gls{dnn} used in \gls{disnets} to compute the non-linear context representation is reported in Table~\ref{tab:network_structure}. 
	Specifically, we consider two convolutional layers of size $(\xi_1,\xi_2,\xi_3)$, where $\xi_1$ is the number of channels, and $\xi_2$ and $\xi_3$ are the kernel width and height, respectively, while the Max Pool field represents the size of the max pooling window. 
	The dimension of the Latent Layer is set to $10$. The size of the Output Layer is equal to the number of feasible single actions, i.e., orthogonal channels, $K$. 
	The variance decaying factor is $\gamma = 0.9999$, and number of steps between two network updates (see Algorithm~\ref{alg:neural_linear_ts}, line $10$) is equal to $O=100$.
	\gls{disnets}' DNN implements the \emph{Leaky Rectified Linear Unit (Leaky ReLu)} activation function.

	\paragraph{Performance metrics} 
	\label{par:performance_metrics}
	
	Numerical results are given in terms of the overhead (measured as the impact of FCI transmissions on the control channel), and the \gls{e2e} latency and reliability defined in Sec.~\ref{subsec:latency}, as a function of the number of UEs and the type of traffic.
	
	\paragraph{Benchmarks} 
	\label{par:benchmarks}
	
	The performance of \gls{disnets} is compared against the following baselines:
	\begin{itemize}
		\item \gls{gbs}: it implements the standard centralized 5G NR \gls{gbs}~\cite{36213}, which requires \glspl{ue} and the \gls{gnb} to exchange scheduling requests (via the \gls{pucch}) and grants (via the \gls{pdcch}), respectively, before transmitting data. In this case resource allocation is based on the number and size of packets that the UEs have in their queues when transmitting scheduling requests via the PUCCH.
		\item \gls{sps}: it implements the standard centralized 5G NR \gls{sps}~\cite{R1-167309}, in which the gNB allocates (part of) the~resources to the UEs semi-statically over a certain time interval. This approach promotes lower latency as resources are assigned only when UEs generate packets, and without additional message exchanges during~transmission.
		\item \gls{nlts}: it implements distributed resource allocation based on the \gls{nlts} algorithm proposed in~\cite{Riquelme2018} and described in Sec.~\ref{sub:nlts}(b), thus with the assumption that UEs can transmit through one single orthogonal channel.
		\item {RandomK}: it implements distributed resource allocation, in which UEs use exactly $K^*$ orthogonal channels (i.e., \glspl{rb}) for each of their packet transmissions, and $K^*$ is optimized via exhaustive~search. Notice that RandomK is not state-of-the-art, but has been explicitly proposed, designed, and implemented to have another decentralized benchmark to compare the performance of DISNETS.
	\end{itemize}

	\subsection{Training Convergence}
	\label{sub:num_traninig}
	
	\begin{figure}[!t]
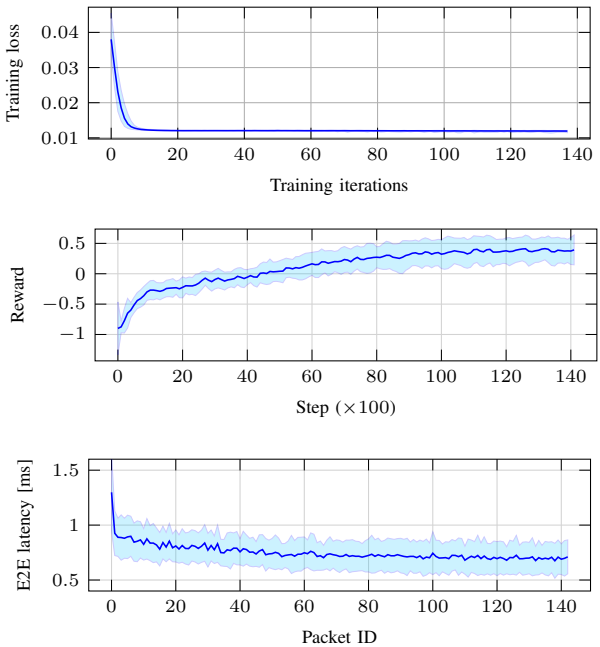

		\centering 
		\begin{subfigure}[t!]{0.99\columnwidth}
			\centering 
			\setlength\fwidth{0.8\columnwidth}
			\setlength\fheight{0.2\columnwidth}
			\input{imgs/correlation_ts_2_60_ue_2_6_training_loss.tex}
			\label{fig:training_loss_curve}
		\end{subfigure}\\\vspace{0.33cm}
		\begin{subfigure}[t!]{0.99\columnwidth}
			\centering 
			\setlength\fwidth{0.8\columnwidth}
			\setlength\fheight{0.2\columnwidth}
			\input{imgs/correlation_ts_2_60_ue_2_6_avg_reward.tex}
			\label{fig:training_reward_curve}
		\end{subfigure}\\\vspace{0.33cm}
		\begin{subfigure}[t!]{0.99\columnwidth}
			\centering 
			\setlength\fwidth{0.8\columnwidth}
			\setlength\fheight{0.2\columnwidth}
			\input{imgs/correlation_ts_2_60_ue_2_6_lat_training.tex}
			\label{fig:e2e_latency_curve}
		\end{subfigure}
		\caption{Convergence performance of \gls{disnets} in terms of empirical average and standard deviation of the training loss (top), reward (center), and \gls{e2e} latency (bottom). We consider uniformly aperiodic traffic, with $t_{min}=2$ ms, $t_{max}=6$ ms, and $N=60$.\vspace{-0.33cm}}
		\label{fig:training}
	\end{figure}
	
	First, we study the training performance of the proposed \gls{disnets} framework considering {uniformly aperiodic} traffic, with $t_{min}=2$ ms, $t_{max} = 6$ ms, and $N=60$ \glspl{ue}.
	Fig.~\ref{fig:training} (top) plots the average and standard deviation of the training loss of the NLTS module  of \gls{disnets} (specifically, the DNN), which is used to learn the non-linear representation of the context $\phi_\omega(s)$.
	The loss decreases quite quickly, and becomes stable after around $0.5$ s, which is an indication of the accuracy of the \gls{disnets} implementation. In fact, as the training progresses, UEs are learning to allocate resources more~accurately.
	
	In Fig.~\ref{fig:training} (center) we plot the statistics of the reward obtained by the agents as a function of the total number of interactions with the environment. 
	We can see that, at the beginning, the reward is close to $-1$ given that all \glspl{ue} start making random decisions in terms of the orthogonal channels to use at each \gls{su}.
	Then, the average reward is an increasing function of the number of steps, meaning that the UEs are learning to make more accurate allocations as the training progresses.
	Interestingly, we can recognize two training phases. 
	At first the UEs are learning fast, at the rate of convergence of the DNN:  in this phase, the average reward increases steeply.
	This is motivated by the fact that, at the beginning of the training phase, many collisions occur, which gives \gls{disnets} more chances to optimize based on the relative rewards. 
	Then, \gls{disnets} takes more time to achieve better cooperation, and the framework optimizes more slowly.
	
	In Fig.~\ref{fig:training} (bottom) we plot the statistics of the \gls{e2e} latency experienced by the UEs as a function of the packet ID. Again, it is possible to separate the two training regimes. At first, \gls{disnets} can estimate the number of orthogonal channels to use to minimize collisions starting from a random guess, and achieves an average latency of around 1 ms. Then, \gls{disnets} is used to further optimize resource allocation reducing the latency to around $0.7$ ms, though taking more time to converge.

\begin{table}[b!]
\renewcommand{\arraystretch}{1.4}
\centering
\footnotesize
\caption{Size (in bits) of the FCI and DCI signals, vs. the number of active UEs ($N_a$) and the number of orthogonal channels ($K$). The total number of UEs in the system is set to $N=500$.}
\label{tab:dci}
\begin{tabular}{|c|c|c|}
\hline
\multirow{2}{*}{FCI} & \multicolumn{2}{c|}{DCI} \\\cline{2-3}
                     & DCI$_m$        & DCI$_M$        \\\hline
 $K\cdot\log_2(N + 2)$ &   $N_a\cdot(\log_2 K + 10)$         &    $N_a\cdot(\log_2 K + 37)$ \\ \hline
\end{tabular}
\end{table}

	\subsection{Overhead} 
	\label{sec:overhead}
	As described in Sec.~\ref{sub:fci_context}, \gls{disnets} requires periodic \gls{fci} transmissions, which include the transmission outcomes (successful transmission, collision, or outage) relative to every UE on each orthogonal channel, and is used to generate the context and the reward. Therefore, the size of the \gls{fci} (in bits) can be computed as $K\cdot\log_2(N + 3)$ (see Table~\ref{tab:dci}).
	
The structure of the FCI is similar to that of the 5G NR \gls{dci} signal~\cite{parkvall20185g}, which is used to handle downlink transmissions. 
The size of the DCI depends on the number of orthogonal channels $K$. 
Moreover, while the FCI embeds information for all UEs, the DCI is transmitted only to the \emph{active} UEs $N_a\subseteq N$, so the overall size (in bits) goes as $N_a\cdot\log_2(K)$.
Furthermore, the DCI requires an additional (variable) number of bits to carry information for, e.g., carrier aggregation, \gls{harq}, frequency allocation, channel access~\cite[Sec. 10.1.4]{parkvall20185g}. 3GPP NR defines 10$+$ different DCI formats. For simplicity, we consider two representative DCI formats, namely DCI$_m$ and DCI$_M$, which require $10$ and $37$ additional bits, respectively, and the relative DCI size is reported in Table~\ref{tab:dci}.

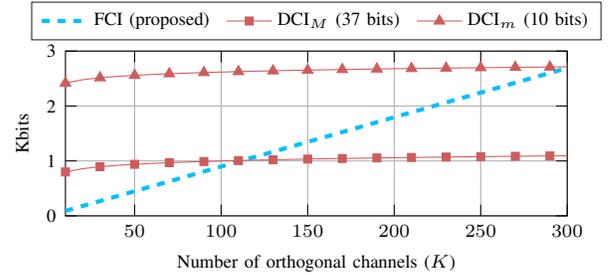
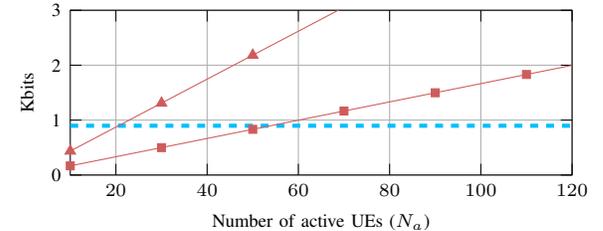
\begin{figure}[!t]
	\centering 
	\begin{subfigure}[t!]{0.99\columnwidth}
	\centering 
	\setlength\fwidth{0.8\columnwidth}
  \setlength\fheight{0.25\columnwidth}
\pgfplotsset{scaled y ticks=false}
\begin{tikzpicture}
\pgfplotsset{
tick label style={font=\scriptsize},
label style={font=\scriptsize},
legend  style={font=\scriptsize}
}

\definecolor{darkgray176}{RGB}{176,176,176}
\definecolor{bluemp}{RGB}{2,191,255}
\definecolor{redmp}{RGB}{205,92,91}
\definecolor{violmp}{RGB}{230,230,250}
\definecolor{orangep}{RGB}{255,159,122,255}

\begin{axis}[
width=0.951\fwidth,
height=\fheight,
at={(0\fwidth,0\fheight)},
scale only axis,
x grid style={darkgray176},
xlabel={Number of orthogonal channels ($K$)},
xmajorgrids,
xmin=10, xmax=300,
xtick style={color=black},
y grid style={darkgray176},
ytick={0,1000,2000,3000,4000},
yticklabels={0,1,2,3,4},
ylabel={Kbits},
yticklabel style={
        /pgf/number format/fixed,
        /pgf/number format/precision=5
},
ymajorgrids,
ymin=0, ymax=3000,
ytick style={color=black},
legend style={legend cell align=left, align=left, draw=white!15!black, at={(0.5,1.3)},/tikz/every even column/.append style={column sep=0.15cm},
  anchor=north ,legend columns=-1},
]

\addplot [color=bluemp, dashed,line width=0.55mm]
table {%
10  89.7441459
20  179.4882918
30  269.2324377
40  358.9765836
50  448.7207295
60  538.4648754
70  628.2090213
80  717.9531672
90  807.6973131
100 897.441459
110 987.1856049
120 1076.929751
130 1166.673897
140 1256.418043
150 1346.162188
160 1435.906334
170 1525.65048
180 1615.394626
190 1705.138772
200 1794.882918
210 1884.627064
220 1974.37121
230 2064.115356
240 2153.859502
250 2243.603647
260 2333.347793
270 2423.091939
280 2512.836085
290 2602.580231
300 2692.324377
};
\addlegendentry{FCI (proposed)}

\addplot [color=redmp,  mark size=1.5pt, mark=square*, mark options={solid, fill=redmp, redmp},mark repeat=2]
table {%
10 799.3156857
20 859.3156857
30 894.4134357
40 919.3156857
50 938.6313714
60 954.4134357
70 967.756981
80 979.3156857
90 989.5111858
100  998.6313714
110  1006.881583
120  1014.413436
130  1021.342069
140  1027.756981
150  1033.729121
160  1039.315686
170  1044.563456
180  1049.511186
190  1054.191336
200  1058.631371
210  1062.854731
220  1066.881583
230  1070.729403
240  1074.413436
250  1077.947057
260  1081.342069
270  1084.608936
280  1087.756981
290  1090.794545
300  1093.729121
};
\addlegendentry{DCI$_M$ (37 bits)}

\addplot [color=redmp,  mark size=2.3pt, mark=triangle*, mark options={solid, fill=redmp, redmp},mark repeat=2]
table {%
10  2419.315686
20  2479.315686
30  2514.413436
40  2539.315686
50  2558.631371
60  2574.413436
70  2587.756981
80  2599.315686
90  2609.511186
100  2618.631371
110  2626.881583
120  2634.413436
130  2641.342069
140  2647.756981
150  2653.729121
160  2659.315686
170  2664.563456
180  2669.511186
190  2674.191336
200  2678.631371
210  2682.854731
220  2686.881583
230  2690.729403
240  2694.413436
250  2697.947057
260  2701.342069
270  2704.608936
280  2707.756981
290  2710.794545
300  2713.729121
};
\addlegendentry{DCI$_m$ (10 bits)}

\end{axis}

\end{tikzpicture}
		\caption{Impact of the number of channels. We set $N_a=60$ and $N=500$.}
		\label{fig:dci-rb}
	\end{subfigure}\\\vspace{0.33cm}
	\begin{subfigure}[t!]{0.99\columnwidth}
	\centering 
	\setlength\fwidth{0.8\columnwidth}
  \setlength\fheight{0.25\columnwidth}
\pgfplotsset{scaled y ticks=false}
\begin{tikzpicture}
\pgfplotsset{
tick label style={font=\scriptsize},
label style={font=\scriptsize},
legend  style={font=\scriptsize}
}

\definecolor{darkgray176}{RGB}{176,176,176}
\definecolor{bluemp}{RGB}{2,191,255}
\definecolor{redmp}{RGB}{205,92,91}
\definecolor{violmp}{RGB}{230,230,250}
\definecolor{orangep}{RGB}{255,159,122,255}

\begin{axis}[
width=0.951\fwidth,
height=\fheight,
at={(0\fwidth,0\fheight)},
scale only axis,
x grid style={darkgray176},
xlabel={Number of active UEs ($N_a$)},
xmajorgrids,
xmin=10, xmax=120,
xtick style={color=black},
y grid style={darkgray176},
ytick={0,1000,2000,3000,4000},
yticklabels={0,1,2,3,4},
ylabel={Kbits},
yticklabel style={
        /pgf/number format/fixed,
        /pgf/number format/precision=5
},
ymajorgrids,
ymin=0, ymax=3000,
ytick style={color=black},
]

\addplot [color=bluemp, dashed,line width=0.55mm]
table {%
10  897.441459
20  897.441459
30  897.441459
40  897.441459
50  897.441459
60  897.441459
70  897.441459
80  897.441459
90  897.441459
100 897.441459
110 897.441459
120 897.441459
};

\addplot [color=redmp,  mark size=1.5pt, mark=square*, mark options={solid, fill=redmp, redmp},mark repeat=2]
table {%
10  166.4385619
20  332.8771238
30  499.3156857
40  665.7542476
50  832.1928095
60  998.6313714
70  1165.069933
80  1331.508495
90  1497.947057
100 1664.385619
110 1830.824181
120 1997.262743
};

\addplot [color=redmp,  mark size=2.3pt, mark=triangle*, mark options={solid, fill=redmp, redmp},mark repeat=2]
table {%
10  436.4385619
20  872.8771238
30  1309.315686
40  1745.754248
50  2182.192809
60  2618.631371
70  3055.069933
80  3491.508495
90  3927.947057
100 4364.385619
110 4800.824181
120 5237.262743
};

\end{axis}

\end{tikzpicture}
		\caption{Impact of the number of active UEs. We set $N=500$ and $K=100$.}
		\label{fig:dci-ue}
	\end{subfigure}
	\caption{Overhead performance measured in terms of the size of the \gls{fci} (proposed) vs. the 3GPP NR \gls{dci}, as a function of the number of orthogonal channels (top) and UEs (bottom). We consider two DCI formats, namely DCI$_m$ and DCI$_M$, which require up to 10 and 37 additional bits, respectively, for resource allocation~\cite{parkvall20185g}.}
	\label{fig:dci}
\end{figure}


\begin{figure}[!t]
	\centering 
	\begin{subfigure}[t!]{0.99\columnwidth}
		\centering 
		\setlength\fwidth{0.8\columnwidth}
		\setlength\fheight{0.25\columnwidth}
\pgfplotsset{scaled y ticks=false}
\begin{tikzpicture}
\pgfplotsset{
tick label style={font=\scriptsize},
label style={font=\scriptsize},
legend  style={font=\scriptsize}
}

\definecolor{darkgray176}{RGB}{176,176,176}
\definecolor{bluemp}{RGB}{2,191,255}
\definecolor{redmp}{RGB}{205,92,91}
\definecolor{violmp}{RGB}{230,230,250}
\definecolor{orangep}{RGB}{255,159,122,255}

\begin{axis}[
width=0.951\fwidth,
height=\fheight,
at={(0\fwidth,0\fheight)},
scale only axis,
x grid style={darkgray176},
xlabel={Number of orthogonal channels ($K$)},
xmajorgrids,
xmin=0, xmax=15,
xtick style={color=black},
y grid style={darkgray176},
ylabel={\acrshort{cdf}},
yticklabel style={
        /pgf/number format/fixed,
        /pgf/number format/precision=5
},
ymajorgrids,
ymin=0, ymax=1,
ytick style={color=black},
legend style={legend cell align=left, align=left, draw=white!15!black, at={(0.5,1.3)},/tikz/every even column/.append style={column sep=0.15cm},
  anchor=north ,legend columns=-1},
]

\addplot [color=bluemp, dashed,line width=0.55mm]
table {%
0  0
1  0.03
2  0.07
3  0.11
4  0.13
5  0.18
6  0.2
7  0.22
8  0.255
9  0.27
10 0.3
11 0.32
12 0.32
13 0.36
14 0.37
15 1
};
\addlegendentry{DISNETS (proposed)}

\addplot [color=redmp,  mark size=1.5pt, mark=square*, mark options={solid, fill=redmp, redmp},mark repeat=2]
table {%
0  0
1  0
2  0
3  0
4  0
5  1
6  1
7  1
8  1
9  1
10 1
11 1
12 1
13 1
14 1
15 1
};
\addlegendentry{RandomK}

\end{axis}

\end{tikzpicture}
		\caption{$40$ UEs.}
		\label{fig:cum_k_40}
	\end{subfigure}\\\vspace{0.33cm}
	\begin{subfigure}[t!]{0.99\columnwidth}
		\centering 
		\setlength\fwidth{0.8\columnwidth}
		\setlength\fheight{0.25\columnwidth}
\pgfplotsset{scaled y ticks=false}
\begin{tikzpicture}
\pgfplotsset{
tick label style={font=\scriptsize},
label style={font=\scriptsize},
legend  style={font=\scriptsize}
}

\definecolor{darkgray176}{RGB}{176,176,176}
\definecolor{bluemp}{RGB}{2,191,255}
\definecolor{redmp}{RGB}{205,92,91}
\definecolor{violmp}{RGB}{230,230,250}
\definecolor{orangep}{RGB}{255,159,122,255}

\begin{axis}[
width=0.951\fwidth,
height=\fheight,
at={(0\fwidth,0\fheight)},
scale only axis,
x grid style={darkgray176},
xlabel={Number of orthogonal channels ($K$)},
xmajorgrids,
xmin=0, xmax=15,
xtick style={color=black},
y grid style={darkgray176},
ylabel={\acrshort{cdf}},
yticklabel style={
        /pgf/number format/fixed,
        /pgf/number format/precision=5
},
ymajorgrids,
ymin=0, ymax=1,
ytick style={color=black},
]

\addplot [color=bluemp, dashed,line width=0.55mm]
table {%
0  0.00
1  0.19
2  0.27
3  0.38
4  0.48
5  0.55
6  0.59
7  0.63
8  0.668
9  0.70
10 0.74
11 0.75
12 0.76
13 0.78
14 0.79
15 1.00
};

\addplot [color=redmp,  mark size=1.5pt, mark=square*, mark options={solid, fill=redmp, redmp},mark repeat=2]
table {%
0  0
1  0
2  1
3  1
4  1
5  1
6  1
7  1
8  1
9  1
10 1
11 1
12 1
13 1
14 1
15 1
};

\end{axis}

\end{tikzpicture}
		\caption{$100$ UEs.}
		\label{fig:cum_k_100}
	\end{subfigure}
	\caption{Empirical \acrshort{cdf} of the number of orthogonal channels used at each scheduling opportunity relative to the last $10$ packets considering \gls{disnets} vs. RandomK, as a function of the number of UEs.\vspace{-0.33cm}}
	\label{fig:ks_ue}
\end{figure}
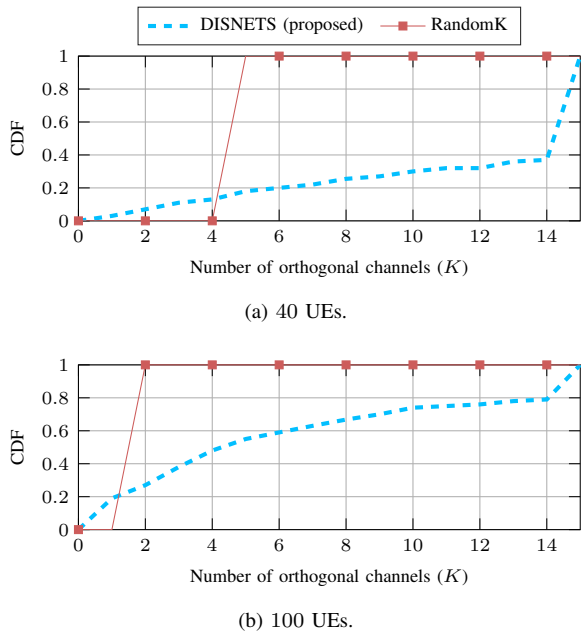

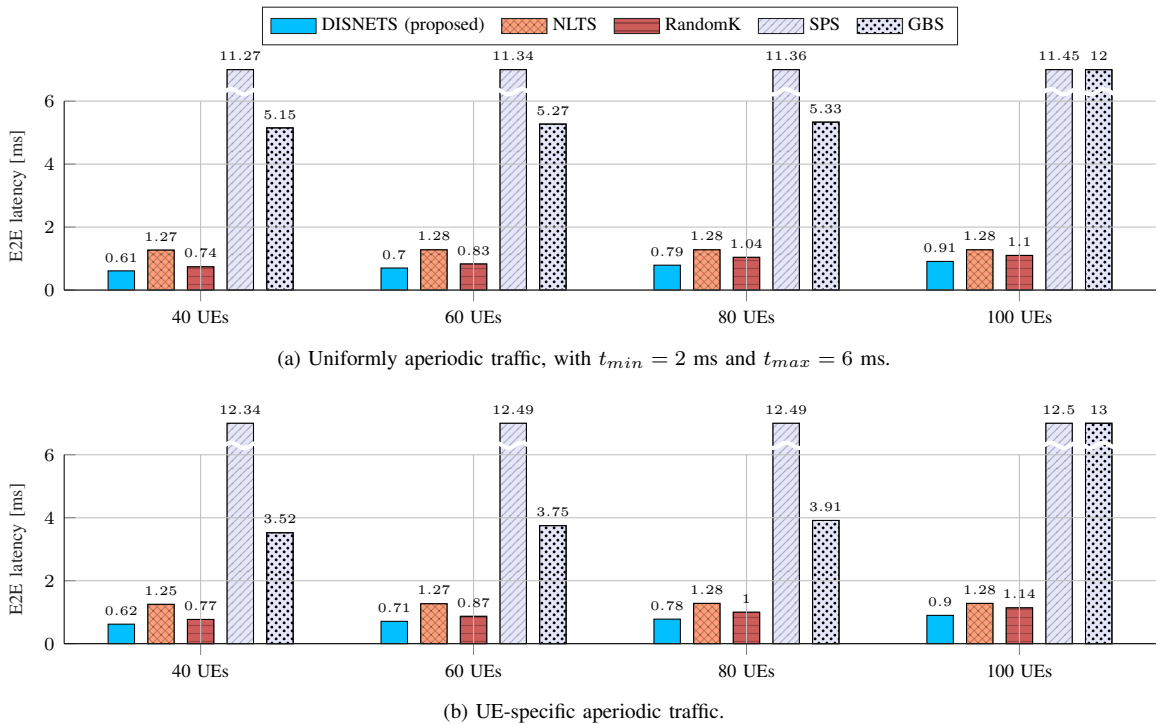
\begin{figure*}[!t]
	\centering 
	\begin{subfigure}[t!]{0.99\textwidth}
		\centering 
		\setlength\fwidth{0.85\textwidth}
		\setlength\fheight{0.14\columnwidth}
\pgfplotsset{scaled y ticks=false}
\pgfplotsset{scaled x ticks=false}
\begin{tikzpicture}
\pgfplotsset{
tick label style={font=\scriptsize},
label style={font=\scriptsize},
legend  style={font=\scriptsize}
}

\definecolor{darkgray176}{RGB}{176,176,176}
\definecolor{bluemp}{RGB}{2,191,255}
\definecolor{redmp}{RGB}{205,92,91}
\definecolor{violmp}{RGB}{230,230,250}
\definecolor{orangep}{RGB}{255,159,122,255}

\begin{axis}[
width=0.951\fwidth,
height=\fheight,
at={(0\fwidth,0\fheight)},
scale only axis,
 every axis plot post/.style={/pgf/number format/fixed},
    ybar=5pt, 
    bar width=10pt,
    xmin=0.5,
	xmax=4.5,
	xtick={1,2,3,4},
	xticklabels={40 UEs,60 UEs,80 UEs,100 UEs},
	xlabel style={font=\scriptsize\color{white!15!black}},
	xticklabel style={align=center},
    ymin=0,
    axis on top,
    ymax=6,
    xmajorgrids,
    ymajorgrids,
    yminorticks=true,
	ylabel style={font=\scriptsize\color{white!15!black}},
	ylabel={\gls{e2e} latency [ms]},
    restrict y to domain*=0:7, 
    visualization depends on=rawy\as\rawy, 
    after end axis/.code={ 
            \draw [ultra thick, white, decoration={snake, amplitude=1pt}, decorate] (rel axis cs:0,1.05) -- (rel axis cs:1,1.05);
        },
    nodes near coords={%
           \tiny\pgfmathprintnumber{\rawy}
        },
    axis lines*=left,
    clip=false,
    legend style={legend cell align=left, align=left, draw=white!15!black, at={(0.5,1.5)},/tikz/every even column/.append style={column sep=0.15cm},
  anchor=north ,legend columns=-1},
    ]
\addplot[ybar, fill=bluemp, area legend] table[row sep=crcr] {%
1 0.61\\  
2 0.70\\
3    0.79\\
4    0.91\\
};
\addlegendentry{\gls{disnets} (proposed)}

\addplot[ybar, fill=orangep, area legend, postaction={pattern=crosshatch, opacity=0.5}] table[row sep=crcr] {%
	1 1.27\\
	2 1.28 \\
	3 1.28 \\
	4 1.28 \\
};
\addlegendentry{\gls{nlts}}

\addplot[ybar, fill=redmp, area legend,postaction={pattern=horizontal lines, opacity=0.5}] table[row sep=crcr] {%
1 0.74\\
2 0.83\\
3 1.04\\
4 1.10\\
};
\addlegendentry{RandomK}

\addplot[ybar, fill=violmp, area legend, postaction={pattern=north east lines, opacity=0.5}] table[row sep=crcr] {%
1 11.2662  \\       
2 11.3367\\ 
3 11.3567 \\
4 11.4528\\
};
\addlegendentry{SPS}

\addplot[ybar, fill=violmp, area legend, postaction={pattern=crosshatch dots}] table[row sep=crcr] {%
1 5.15\\
2 5.27165\\
3 5.3322\\
4 12\\
};
\addlegendentry{GBS}

\end{axis}

\end{tikzpicture}
		\caption{Uniformly aperiodic traffic, with $t_{min}=2$ ms and $t_{max}=6$ ms.}
		\label{fig:comparison_ue_unif}
	\end{subfigure}\\\vspace{0.33cm}
	\begin{subfigure}[t!]{0.99\textwidth}
		\centering 
		\setlength\fwidth{0.85\textwidth}
		\setlength\fheight{0.14\columnwidth}
\pgfplotsset{scaled y ticks=false}
\pgfplotsset{scaled x ticks=false}
\begin{tikzpicture}
\pgfplotsset{
tick label style={font=\scriptsize},
label style={font=\scriptsize},
legend  style={font=\scriptsize}
}

\definecolor{darkgray176}{RGB}{176,176,176}
\definecolor{bluemp}{RGB}{2,191,255}
\definecolor{redmp}{RGB}{205,92,91}
\definecolor{violmp}{RGB}{230,230,250}
\definecolor{orangep}{RGB}{255,159,122,255}

\begin{axis}[
width=0.951\fwidth,
height=\fheight,
at={(0\fwidth,0\fheight)},
scale only axis,
 every axis plot post/.style={/pgf/number format/fixed},
    ybar=5pt,
    bar width=10pt,
    xmin=0.5,
	xmax=4.5,
	xtick={1,2,3,4},
	xticklabels={40 UEs,60 UEs,80 UEs,100 UEs},
	xlabel style={font=\scriptsize\color{white!15!black}},
	xticklabel style={align=center},
    ymin=0,
    axis on top,
    ymax=6,
    xmajorgrids,
    ymajorgrids,
    yminorticks=true,
	ylabel style={font=\scriptsize\color{white!15!black}},
	ylabel={\gls{e2e} latency [ms]},
    restrict y to domain*=0:7, 
    visualization depends on=rawy\as\rawy, 
    after end axis/.code={ 
            \draw [ultra thick, white, decoration={snake, amplitude=1pt}, decorate] (rel axis cs:0,1.05) -- (rel axis cs:1,1.05);
        },
    nodes near coords={%
           \tiny\pgfmathprintnumber{\rawy}
        },
    axis lines*=left,
    clip=false,
    ]
\addplot[ybar, fill=bluemp, area legend] table[row sep=crcr] {%
1 0.62\\
2 0.71\\
3 0.78\\
4 0.9\\
};

\addplot[ybar, fill=orangep, area legend, postaction={pattern=crosshatch, opacity=0.5}] table[row sep=crcr] {%
	1 1.25\\
	2 1.27 \\
	3 1.28 \\
	4 1.28 \\
};

\addplot[ybar, fill=redmp, area legend,postaction={pattern=horizontal lines, opacity=0.5}] table[row sep=crcr] {%
1 0.77  \\
2 0.87\\
3 1\\
4 1.14\\
};

\addplot[ybar, fill=violmp, area legend, postaction={pattern=north east lines, opacity=0.5}] table[row sep=crcr] {%
1 12.3373  \\       
2 12.49185\\ 
3 12.48935 \\
4 12.5\\
};

\addplot[ybar, fill=violmp, area legend, postaction={pattern=crosshatch dots}] table[row sep=crcr] {%
1 3.5225\\
2 3.7495\\
3 3.91315\\
4 13\\
};

\end{axis}

\end{tikzpicture}
		\caption{UE-specific aperiodic traffic.}
		\label{fig:comparison_ue_ue}
	\end{subfigure}
	\caption{Average \gls{e2e} latency for \gls{disnets} vs. RandomK, \gls{nlts}, SPS, and GBS, as a function of the number of UEs and the type of traffic.\vspace{-0.33cm}}
	\label{fig:perf_ue}
\end{figure*}

Based on the above introduction, in Fig.~\ref{fig:dci} we plot the size of the FCI, DCI$_m$, and DCI$_M$ signals, which is directly proportional to the overhead.
Specifically, we see that the size of the FCI scales linearly with the number of orthogonal channels (Fig.~\ref{fig:dci-rb}) and logarithmically with the number of active UEs (Fig.~\ref{fig:dci-ue}), while for the DCI it is almost the opposite. As such, \gls{disnets} achieves comparable or lower overhead than other solutions based on the DCI (e.g., 5G NR \gls{gbs}) in many reasonable configurations.
Notice that the results in Fig.~\ref{fig:dci} do not account for the additional overhead introduced to send (receive) scheduling requests (grants) in \gls{gbs}, which is not required in \gls{disnets} since resource allocation is distributed.

As another measure of overhead, in Fig.~\ref{fig:ks_ue} we plot the empirical \gls{cdf} of the number of orthogonal channels used at each scheduling opportunity for \gls{disnets} vs. RandomK, which is an indication of the channel occupancy.  Statistics are referred to the last 10 packets, i.e., after the convergence of \gls{disnets}. 
We observe that RandomK uses a constant (though optimized) number of channels per UE equal to $K^* \in \{ 5, 2\}$ for $N\in\{40,100\}$ respectively, so $K^*$ is a decreasing function of $N$.  This is due to the fact that, as the number of UEs increases, the number of collisions also increases: in these conditions, the system is encouraged to reduce the number of channels to reduce the probability of collison. 
On the other hand, the adaptability and flexibility features of \gls{disnets} make the number of channels to use for transmission vary significantly; as such, \glspl{ue} are free to optimize the number of resources as a function of $N$.
Notice that, even though RandomK uses on average less resources that DISNETS, it results in more collisions (see Figs. \ref{fig:perf_ue} and
\ref{fig:freq_ue}), which would increase the number of re-transmissions and, eventually, the overall number of resources. Eventually, the performance of RandomK (in terms of latency and reliability) is worse than that of DISNETS because RandomK is unable to~optimize.

\subsection{Performance Evaluation} 
\label{sub:performance_evaluation}


\paragraph{Impact of the number of \glspl{ue} and the resource allocation strategy}


We now compare the performance of \gls{disnets} against those of RandomK, \gls{nlts}, \gls{gbs}, and \gls{sps} baselines as a function of the number of \glspl{ue} in the system.
Fig.~\ref{fig:comparison_ue_unif} reports the \gls{e2e} latency considering uniformly aperiodic traffic. As expected, the \gls{e2e} latency increases as $N$ increases given that the network is more congested, which increases the probability of collision and re-transmissions.
Also, we can see that \gls{disnets} always outperforms all the benchmarks.
In particular, centralized {GBS} is not able to satisfy the $L_{\rm th}=1$~ms requirement of URLLC due to the additional delays introduced to send (receive) scheduling requests (grants), especially when the number of \glspl{ue} increases.
Interestingly, GBS outperforms SPS in case of aperiodic traffic. 
In fact, SPS is designed to work well as long as the traffic is periodic~\cite{cuozzo2022enabling}: in this case, SPS can pre-allocate resources based on the actual traffic periodicity, and does not require the UEs and the gNB to exchange additional messages. However, for aperiodic traffic as in Fig.~\ref{fig:perf_ue}, SPS may not be able to react to possible (unpredictable) changes in the traffic patterns and requests, with respect to how resources were originally pre-assigned, which implies that the system may operate in a sub-optimal configuration~\cite{cavallero2023new}. 
As such, unscheduled UEs will keep data packets in the queue, thus accumulating delays, at least until SPS is re-configured by another \gls{rrc} interaction. 

Compared to another distributed benchmark such as RandomK, \gls{disnets} can reduce the \gls{e2e} latency by up to $20\%$. In fact, \gls{disnets} exploits coordination, and is designed to optimize resource allocation depending on the type of traffic and the relative load of machines (for example allocating more resources to machines with more UEs). 
In the end, the latency for \gls{nlts} is up to $1.71\times $ and $2\times$ higher than \gls{disnets} and RandomK, respectively, given that \glspl{ue} can use only one orthogonal channel. While this approach brings the probability of collision to almost zero, it leaves the network underutilized. In comparison, RandomK uses $K^*$ channels, while \gls{disnets} can dynamically adapt the number of channels to optimize the trade-off between latency and collision.

Moreover, in Fig.~\ref{fig:comparison_ue_ue} we consider the case of {UE-specific aperiodic} traffic. We observe that \gls{disnets} and GBS can exploit the additional degrees of correlation introduced in the traffic to improve the latency compared to Fig.~\ref{fig:comparison_ue_unif}. This is particularly true for $N\leq80$, while for $N>80$ the performance degrades quickly due to congestion.
Still, \gls{disnets} is the only scheme able to satisfy the $L_{\text{th}}=1$ ms latency requirement in all configurations as it learns more from the correlation in the packet generation process. 
Notice that the latency for RandomK and \gls{sps} is slightly higher than in the scenario in Fig.~\ref{fig:comparison_ue_unif} due to the fact that the extra packets generated in the interval [$t_{\rm min}$, $t_{\rm max}$] may create more collisions. 
A similar observation holds for the \gls{nlts} baseline.

\begin{figure*}[!t]
	\centering 
	\begin{subfigure}[t!]{0.99\textwidth}
		\centering 
		\setlength\fwidth{0.8\columnwidth}
		\setlength\fheight{0.45\columnwidth}
\pgfplotsset{scaled y ticks=false}
\pgfplotsset{scaled x ticks=false}
\begin{tikzpicture}
\pgfplotsset{
tick label style={font=\scriptsize},
label style={font=\scriptsize},
legend  style={font=\scriptsize}
}

\definecolor{darkgray176}{RGB}{176,176,176}
\definecolor{bluemp}{RGB}{2,191,255}
\definecolor{redmp}{RGB}{205,92,91}
\definecolor{violmp}{RGB}{230,230,250}
\definecolor{orangep}{RGB}{255,159,122,255}

\begin{axis}[
width=0,
height=0,
at={(0,0)},
scale only axis,
xmin=0,
xmax=0,
xtick={},
ymin=0,
ymax=0,
ytick={},
    legend style={legend cell align=left, align=left, draw=white!15!black, at={(0.6,1.3)},/tikz/every even column/.append style={column sep=0.15cm},
  anchor=north ,legend columns=-1},
    ]

     \addplot[ybar, fill=bluemp, area legend] table[row sep=crcr] {%
0 0 \\
 };
 \addlegendentry{\gls{disnets} (proposed)}

\addplot[ybar, fill=redmp, area legend,postaction={pattern=horizontal lines, opacity=0.5}] table[row sep=crcr] {%
0 0 \\
};
\addlegendentry{RandomK}


\end{axis}

\end{tikzpicture}
	\end{subfigure}\\\vspace{0.2cm}
	\begin{subfigure}[t!]{0.67\columnwidth}
		\centering 
		\setlength\fwidth{0.8\columnwidth}
		\setlength\fheight{0.3\columnwidth}
\pgfplotsset{scaled y ticks=false}
\pgfplotsset{scaled x ticks=false}
\begin{tikzpicture}
\pgfplotsset{
tick label style={font=\scriptsize},
label style={font=\scriptsize},
legend  style={font=\scriptsize}
}

\definecolor{darkgray176}{RGB}{176,176,176}
\definecolor{bluemp}{RGB}{2,191,255}
\definecolor{redmp}{RGB}{205,92,91}
\definecolor{violmp}{RGB}{230,230,250}
\definecolor{orangep}{RGB}{255,159,122,255}

\begin{axis}[
width=0.951\fwidth,
height=\fheight,
at={(0\fwidth,0\fheight)},
scale only axis,
 every axis plot post/.style={/pgf/number format/fixed},
    ybar,
    bar width=4pt,
    xmin=0,
	xmax=1.68,
	xlabel={\gls{e2e} latency [ms]},
	xlabel style={font=\scriptsize\color{white!15!black}},
	xticklabel style={align=center},
    ymin=0,
    axis on top,
    ymax=0.4,
    xmajorgrids,
    ymajorgrids,
    yminorticks=true,
	ylabel style={font=\scriptsize\color{white!15!black}},
	ylabel={\acrshort{pdf}},
    axis lines*=left,
    clip=false,
    ]

     \addplot[ybar, fill=bluemp, area legend] table[row sep=crcr] {%
0.25    0.00 \\
0.33    0.00 \\
0.40    0.01 \\
0.48    0.34 \\
0.55    0.29 \\
0.63    0.16 \\
0.70    0.10 \\
0.78    0.06 \\
0.85    0.04 \\
0.93    0.01 \\
1.00    0.00 \\
1.08    0.01 \\
1.15    0.00 \\
1.23    0.00 \\
1.30    0.00 \\
1.38    0.00 \\
1.45    0.00 \\
1.53    0.00 \\
1.60    0.00 \\
1.68    0.00 \\
 };

\addplot[ybar, fill=redmp, area legend,postaction={pattern=horizontal lines, opacity=0.5}] table[row sep=crcr] {%
0.25   0.00 \\
0.33   0.00 \\
0.40   0.00 \\
0.48   0.01944444 \\
0.55   0.0777777 \\
0.63   0.22 \\
0.70   0.24 \\
0.78   0.19 \\
0.85   0.09 \\
0.93   0.09 \\
1.00   0.04 \\
1.08   0.03 \\
1.15   0.00 \\
1.23   0.00 \\
1.30   0.00 \\
1.38   0.00 \\
1.45   0.00 \\
1.53   0.00 \\
1.60   0.00 \\
1.68   0.00 \\
};


\end{axis}

\end{tikzpicture}
	\end{subfigure}
	\begin{subfigure}[t!]{0.67\columnwidth}
		\centering 
		\setlength\fwidth{0.8\columnwidth}
		\setlength\fheight{0.3\columnwidth}
\pgfplotsset{scaled y ticks=false}
\pgfplotsset{scaled x ticks=false}
\begin{tikzpicture}
\pgfplotsset{
tick label style={font=\scriptsize},
label style={font=\scriptsize},
legend  style={font=\scriptsize}
}

\definecolor{darkgray176}{RGB}{176,176,176}
\definecolor{bluemp}{RGB}{2,191,255}
\definecolor{redmp}{RGB}{205,92,91}
\definecolor{violmp}{RGB}{230,230,250}
\definecolor{orangep}{RGB}{255,159,122,255}

\begin{axis}[
width=0.951\fwidth,
height=\fheight,
at={(0\fwidth,0\fheight)},
scale only axis,
 every axis plot post/.style={/pgf/number format/fixed},
    ybar,
    bar width=4pt,
    xmin=0,
	xmax=1.68,
	xlabel={\gls{e2e} latency [ms]},
	xlabel style={font=\scriptsize\color{white!15!black}},
	xticklabel style={align=center},
    ymin=0,
    axis on top,
    ymax=0.4,
    xmajorgrids,
    ymajorgrids,
    yminorticks=true,
	ylabel style={font=\scriptsize\color{white!15!black}},
	ylabel={\acrshort{pdf}},
    axis lines*=left,
    clip=false,
    ]

     \addplot[ybar, fill=bluemp, area legend] table[row sep=crcr] {%
0.25    0.00 \\
0.33    0.00 \\
0.40    0.00 \\
0.48    0.14 \\
0.55    0.24 \\
0.63    0.20 \\
0.70    0.14 \\
0.78    0.10 \\
0.85    0.08 \\
0.93    0.03 \\
1.00    0.03 \\
1.08    0.01 \\
1.15    0.01 \\
1.23    0.01 \\
1.30    0.01 \\
1.38    0.00 \\
1.45    0.00 \\
1.53    0.00 \\
1.60    0.00 \\
1.68    0.00 \\
 };

\addplot[ybar, fill=redmp, area legend,postaction={pattern=horizontal lines, opacity=0.5}] table[row sep=crcr] {%
0.25   0.00 \\
0.33   0.00 \\
0.40   0.00 \\
0.48   0 \\
0.55   0.33333333 \\
0.63   0.15 \\
0.70   0.13 \\
0.78   0.19 \\
0.85   0.15 \\
0.93   0.12 \\
1.00   0.09 \\
1.08   0.07 \\
1.15   0.04 \\
1.23   0.02 \\
1.30   0.00 \\
1.38   0.00 \\
1.45   0.00 \\
1.53   0.00 \\
1.60   0.00 \\
1.68   0.00 \\
};


\end{axis}

\end{tikzpicture}
	\end{subfigure}
	\begin{subfigure}[t!]{0.67\columnwidth}
		\centering 
		\setlength\fwidth{0.8\columnwidth}
		\setlength\fheight{0.3\columnwidth}
\pgfplotsset{scaled y ticks=false}
\pgfplotsset{scaled x ticks=false}
\begin{tikzpicture}
\pgfplotsset{
tick label style={font=\scriptsize},
label style={font=\scriptsize},
legend  style={font=\scriptsize}
}

\definecolor{darkgray176}{RGB}{176,176,176}
\definecolor{bluemp}{RGB}{2,191,255}
\definecolor{redmp}{RGB}{205,92,91}
\definecolor{violmp}{RGB}{230,230,250}
\definecolor{orangep}{RGB}{255,159,122,255}

\begin{axis}[
width=0.951\fwidth,
height=\fheight,
at={(0\fwidth,0\fheight)},
scale only axis,
 every axis plot post/.style={/pgf/number format/fixed},
    ybar,
    bar width=4pt,
    xmin=0,
	xmax=1.68,
	xlabel={\gls{e2e} latency [ms]},
	xlabel style={font=\scriptsize\color{white!15!black}},
	xticklabel style={align=center},
    ymin=0,
    axis on top,
    ymax=0.4,
    xmajorgrids,
    ymajorgrids,
    yminorticks=true,
	ylabel style={font=\scriptsize\color{white!15!black}},
	ylabel={\acrshort{pdf}},
    axis lines*=left,
    clip=false,
    ]

     \addplot[ybar, fill=bluemp, area legend] table[row sep=crcr] {%
0.25    0.00 \\
0.33    0.00 \\
0.40    0.00 \\
0.48    0.01 \\
0.55    0.08 \\
0.63    0.12 \\
0.70    0.16 \\
0.78    0.12 \\
0.85    0.12 \\
0.93    0.09 \\
1.00    0.07 \\
1.08    0.06 \\
1.15    0.05 \\
1.23    0.04 \\
1.30    0.01 \\
1.38    0.02309237 \\
1.45    0.02 \\
1.53    0.01 \\
1.60    0.00 \\
1.68    0.00 \\
 };

\addplot[ybar, fill=redmp, area legend,postaction={pattern=horizontal lines, opacity=0.5}] table[row sep=crcr] {%
0.25   0.00 \\
0.33   0.00 \\
0.40   0.00 \\
0.48   0 \\
0.55   0.01559252 \\
0.63   0.04 \\
0.70   0.05 \\
0.78   0.10 \\
0.85   0.07 \\
0.93   0.09 \\
1.00   0.10 \\
1.08   0.11 \\
1.15   0.11 \\
1.23   0.06340956 \\
1.30   0.07 \\
1.38   0.06 \\
1.45   0.04 \\
1.53   0.03 \\
1.60   0.02 \\
1.68   0.02 \\
};


\end{axis}

\end{tikzpicture}
	\end{subfigure}\\\vspace{0.33cm}
	\begin{subfigure}[t!]{0.99\textwidth}
		\centering 
		\setlength\fwidth{0.8\columnwidth}
		\setlength\fheight{0.45\columnwidth}
\pgfplotsset{scaled y ticks=false}
\pgfplotsset{scaled x ticks=false}
\begin{tikzpicture}
\pgfplotsset{
tick label style={font=\scriptsize},
label style={font=\scriptsize},
legend  style={font=\scriptsize}
}

\definecolor{darkgray176}{RGB}{176,176,176}
\definecolor{bluemp}{RGB}{2,191,255}
\definecolor{redmp}{RGB}{205,92,91}
\definecolor{violmp}{RGB}{230,230,250}
\definecolor{orangep}{RGB}{255,159,122,255}

\begin{axis}[
width=0,
height=0,
at={(0,0)},
scale only axis,
xmin=0,
xmax=0,
xtick={},
ymin=0,
ymax=0,
ytick={},
    legend style={legend cell align=left, align=left, draw=white!15!black, at={(0.6,1.3)},/tikz/every even column/.append style={column sep=0.15cm},
  anchor=north ,legend columns=-1},
    ]

\addplot [color=bluemp, dashed,line width=0.55mm]
table {%
-1 0
};
\addlegendentry{\gls{disnets} (proposed)}

\addplot [color=redmp,  mark size=1.5pt, mark=square*, mark options={solid, fill=redmp, redmp},mark repeat=2]
table {%
-1 0
};
\addlegendentry{RandomK}

\draw [thick,dash dot] (1,0) -- (1,1);

\addplot [thick,dash dot]
table {%
-1 0
};
\addlegendentry{URLLC requirement}


\end{axis}

\end{tikzpicture}
	\end{subfigure}\\\vspace{0.2cm}
	\begin{subfigure}[t!]{0.67\columnwidth}
		\centering 
		\setlength\fwidth{0.8\columnwidth}
		\setlength\fheight{0.4\columnwidth}
\pgfplotsset{scaled y ticks=false}
\begin{tikzpicture}
\pgfplotsset{
tick label style={font=\scriptsize},
label style={font=\scriptsize},
legend  style={font=\scriptsize}
}

\definecolor{darkgray176}{RGB}{176,176,176}
\definecolor{bluemp}{RGB}{2,191,255}
\definecolor{redmp}{RGB}{205,92,91}
\definecolor{violmp}{RGB}{230,230,250}
\definecolor{orangep}{RGB}{255,159,122,255}

\begin{axis}[
width=0.951\fwidth,
height=\fheight,
at={(0\fwidth,0\fheight)},
scale only axis,
x grid style={darkgray176},
xlabel={\gls{e2e} latency [ms]},
xmajorgrids,
xmin=0, xmax=1.75,
xtick style={color=black},
y grid style={darkgray176},
ylabel={\acrshort{cdf}},
yticklabel style={
        /pgf/number format/fixed,
        /pgf/number format/precision=5
},
ymajorgrids,
clip=false,
ymin=0, ymax=1,
ytick style={color=black},
]

\addplot [color=bluemp, dashed,line width=0.55mm]
table {%
0.25		0.00
0.33		0.00
0.40		0.01
0.48		0.3475
0.55		0.63
0.63		0.79
0.70		0.89
0.78		0.95
0.85		0.99
0.93		1.00
1.00		1.00
1.08		1.00
1.15		1.00
1.23		1.00
1.30		1.00
1.38		1.00
1.45		1.00
1.53		1.00
1.60		1.00
1.68		1.00
};

\addplot [color=redmp,  mark size=1.5pt, mark=square*, mark options={solid, fill=redmp, redmp},mark repeat=2]
table {%
0.25	0.00
0.33	0.00
0.40	0.00
0.48	0.01944444
0.55	0.097222222
0.63	0.31
0.70	0.56
0.78	0.75
0.85	0.84
0.93	0.94
1.00	0.98
1.08	1.00
1.15	1.00
1.23	1.00
1.30	1.00
1.38	1.00
1.45	1.00
1.53	1.00
1.60	1.00
1.68	1.00	
};

 \draw [thick,dash dot] (1,0) -- (1,1);

 \draw [thick,dash dot] (0,1) -- (1,1);
 \draw [thick,dash dot] (0,0.98) -- (1,0.98);

 \draw (0.15,1) node[anchor=north] {\tiny $0.98$};
 \draw (1,0.2) node[rotate=90,anchor=north,align=left] {\tiny  URLLC \\[-1.5ex] \tiny requirement};

\end{axis}

\end{tikzpicture}
		\caption{$40$ UEs.}
		\label{fig:comparison_traffic_uniff_40}
	\end{subfigure}
	\begin{subfigure}[t!]{0.67\columnwidth}
		\centering 
		\setlength\fwidth{0.8\columnwidth}
		\setlength\fheight{0.4\columnwidth}
\pgfplotsset{scaled y ticks=false}
\begin{tikzpicture}
\pgfplotsset{
tick label style={font=\scriptsize},
label style={font=\scriptsize},
legend  style={font=\scriptsize}
}

\definecolor{darkgray176}{RGB}{176,176,176}
\definecolor{bluemp}{RGB}{2,191,255}
\definecolor{redmp}{RGB}{205,92,91}
\definecolor{violmp}{RGB}{230,230,250}
\definecolor{orangep}{RGB}{255,159,122,255}

\begin{axis}[
width=0.951\fwidth,
height=\fheight,
at={(0\fwidth,0\fheight)},
scale only axis,
x grid style={darkgray176},
xlabel={\gls{e2e} latency [ms]},
xmajorgrids,
xmin=0, xmax=1.75,
xtick style={color=black},
y grid style={darkgray176},
ylabel={\acrshort{cdf}},
yticklabel style={
        /pgf/number format/fixed,
        /pgf/number format/precision=5
},
ymajorgrids,
ymin=0, ymax=1,
ytick style={color=black},
]

\addplot [color=bluemp, dashed,line width=0.55mm]
table {%
0.25	0.00	
0.33	0.00	
0.40	0.00	
0.48	0.14	
0.55	0.38	
0.63	0.59	
0.70	0.73	
0.78	0.83	
0.85	0.90	
0.93	0.93	
1.00	0.96	
1.08	0.97	
1.15	0.98	
1.23	0.99	
1.30	1.00	
1.38	1.00	
1.45	1.00	
1.53	1.00	
1.60	1.00	
1.68	1.00	
};

\addplot [color=redmp,  mark size=1.5pt, mark=square*, mark options={solid, fill=redmp, redmp},mark repeat=2]
table {%
0.25	0.00
0.33	0.00
0.40	0.00
0.48	0
0.55	0.0333333
0.63	0.18
0.70	0.31
0.78	0.50
0.85	0.65
0.93	0.78
1.00	0.87
1.08	0.94
1.15	0.98
1.23	1.00
1.30	1.00
1.38	1.00
1.45	1.00
1.53	1.00
1.60	1.00
1.68	1.00
};

 \draw [thick,dash dot] (1,0) -- (1,1);

 \draw [thick,dash dot] (0,0.96) -- (1,0.96);
 \draw [thick,dash dot] (0,0.87) -- (1,0.87);

 \draw (0.15,0.98) node[anchor=north] {\tiny $0.96$};
 \draw (0.15,0.89) node[anchor=north] {\tiny $0.87$};
 \draw (1,0.2) node[rotate=90,anchor=north,align=left] {\tiny  URLLC \\[-1.5ex] \tiny requirement};

\end{axis}

\end{tikzpicture}
		\caption{$60$ UEs.}
	\end{subfigure}
	\begin{subfigure}[t!]{0.67\columnwidth}
		\centering 
		\setlength\fwidth{0.8\columnwidth}
		\setlength\fheight{0.4\columnwidth}
\pgfplotsset{scaled y ticks=false}
\begin{tikzpicture}
\pgfplotsset{
tick label style={font=\scriptsize},
label style={font=\scriptsize},
legend  style={font=\scriptsize}
}

\definecolor{darkgray176}{RGB}{176,176,176}
\definecolor{bluemp}{RGB}{2,191,255}
\definecolor{redmp}{RGB}{205,92,91}
\definecolor{violmp}{RGB}{230,230,250}
\definecolor{orangep}{RGB}{255,159,122,255}

\begin{axis}[
width=0.951\fwidth,
height=\fheight,
at={(0\fwidth,0\fheight)},
scale only axis,
x grid style={darkgray176},
xlabel={\gls{e2e} latency [ms]},
xmajorgrids,
xmin=0, xmax=1.75,
xtick style={color=black},
y grid style={darkgray176},
ylabel={\acrshort{cdf}},
yticklabel style={
        /pgf/number format/fixed,
        /pgf/number format/precision=5
},
ymajorgrids,
ymin=0, ymax=1,
ytick style={color=black},
]

\addplot [color=bluemp, dashed,line width=0.55mm]
table {%
0.25	0.00
0.33	0.00
0.40	0.00
0.48	0.02
0.55	0.10
0.63	0.22
0.70	0.38
0.78	0.50
0.85	0.62
0.93	0.71
1.00	0.79
1.08	0.85
1.15	0.90
1.23	0.94
1.30	0.96
1.38	0.98
1.45	0.99
1.53	0.99
1.60	1.00
1.68	1.00
};

\addplot [color=redmp,  mark size=1.5pt, mark=square*, mark options={solid, fill=redmp, redmp},mark repeat=2]
table {%
0.25	0.00
0.33	0.00
0.40	0.00
0.48	0
0.55	0.01559252
0.63	0.06
0.70	0.11
0.78	0.21
0.85	0.27
0.93	0.36
1.00	0.46
1.08	0.57
1.15	0.68
1.23	0.75
1.30	0.82
1.38	0.88
1.45	0.92
1.53	0.96
1.60	0.98
1.68	1.00
};

 \draw [thick,dash dot] (1,0) -- (1,1);

 \draw [thick,dash dot] (0,0.8) -- (1,0.8);
 \draw [thick,dash dot] (0,0.46) -- (1,0.46);

 \draw (0.15,0.82) node[anchor=north] {\tiny $0.8$};
 \draw (0.15,0.48) node[anchor=north] {\tiny $0.46$};
 \draw (1,0.2) node[rotate=90,anchor=north,align=left] {\tiny  URLLC \\[-1.5ex] \tiny requirement};

\end{axis}

\end{tikzpicture}
		\caption{$100$ UEs.}
		\label{fig:comparison_traffic_uniff}
	\end{subfigure}
	\caption{Empirical \acrshort{pdf} (top) and \acrshort{cdf} (bottom) of the \gls{e2e} latency considering DISNETS vs. RandomK as a function of the number of UEs. We consider uniformly aperiodic traffic, with $t_{min}=2$ ms and $t_{max}=6$ ms.}
	\label{fig:freq_ue}
\end{figure*}
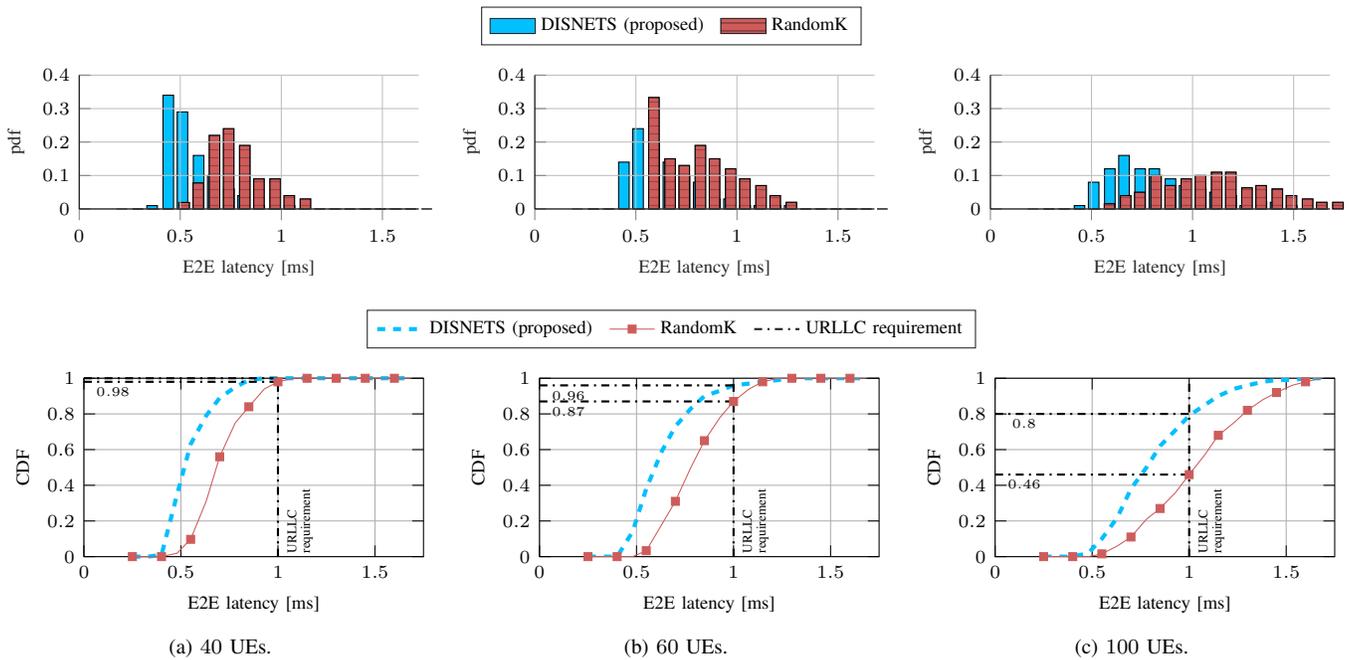

We recall that \gls{urllc} requires both low latency and high reliability.
In Sec.~\ref{subsec:latency} we defined {reliability} $\eta_t(L_{\text{th}})$ as the probability that the \gls{e2e} latency associated with a packet is below a pre-defined requirement (here set to $L_{\rm th}=1$ ms). To capture this trend, in Fig.~\ref{fig:freq_ue} we plot the \gls{pdf} (top) and \gls{cdf} (bottom) of the \gls{e2e} latency for \gls{disnets} vs. {RandomK} (the two best solutions for resource allocation based on the previous results).
We can see that the latency distributions for \gls{disnets} are strongly shifted towards the left compared to RandomK (an indication of a smaller \gls{e2e} latency), and the gap increases as $N$ increases.
For example, while for $N=40$ (Fig.~\ref{fig:comparison_traffic_uniff_40}) both systems achieve comparable performance, for $N=100$ (Fig.~\ref{fig:comparison_traffic_uniff}) we have that only $46\%$ of the UEs experience an \gls{e2e} lower than $L_{\rm th}=1$ ms using RandomK, vs. $80\%$ for \gls{disnets}.

\paragraph{Impact of the type of traffic}
\begin{figure}[!t]
	\centering 
	\begin{subfigure}[t!]{0.99\columnwidth}
		\centering 
		\setlength\fwidth{0.8\columnwidth}
		\setlength\fheight{0.25\columnwidth}
\pgfplotsset{scaled y ticks=false}
\pgfplotsset{scaled x ticks=false}
\begin{tikzpicture}
\pgfplotsset{
tick label style={font=\scriptsize},
label style={font=\scriptsize},
legend  style={font=\scriptsize}
}

\definecolor{darkgray176}{RGB}{176,176,176}
\definecolor{bluemp}{RGB}{2,191,255}
\definecolor{redmp}{RGB}{205,92,91}
\definecolor{violmp}{RGB}{230,230,250}
\definecolor{orangep}{RGB}{255,159,122,255}

\begin{axis}[
width=0.951\fwidth,
height=\fheight,
at={(0\fwidth,0\fheight)},
scale only axis,
 every axis plot post/.style={/pgf/number format/fixed},
    ybar=5pt, 
    bar width=8pt,
    xmin=0.5,
	xmax=4.5,
	xtick={1,2,3,4},
	xticklabels={1,2,3,5},
	xlabel={$t_{min}$ [ms]},
	xlabel style={font=\scriptsize\color{white!15!black}},
	xticklabel style={align=center},
    ymin=0,
    axis on top,
    ymax=2,
    xmajorgrids,
    ymajorgrids,
    yminorticks=true,
	ylabel style={font=\scriptsize\color{white!15!black}},
	ylabel={\gls{e2e} latency [ms]},
    restrict y to domain*=0:2.5, 
    visualization depends on=rawy\as\rawy, 
    after end axis/.code={ 
            \draw [ultra thick, white, decoration={snake, amplitude=1pt}, decorate] (rel axis cs:0,1.08) -- (rel axis cs:1,1.08);
        },
    nodes near coords={%
           \tiny\pgfmathprintnumber{\rawy}
        },
    axis lines*=left,
    clip=false,
    legend style={legend cell align=left, align=left, draw=white!15!black, at={(0.5,1.7)},/tikz/every even column/.append style={column sep=0.15cm},
  anchor=north ,legend columns=-1},
    ]
\addplot[ybar, fill=bluemp, area legend] table[row sep=crcr] {%
1 0.70    \\
2 0.70    \\
3 0.71    \\
4 0.75\\
};
\addlegendentry{\gls{disnets} (proposed)}

\addplot[ybar, fill=redmp, area legend,postaction={pattern=horizontal lines, opacity=0.5}] table[row sep=crcr] {%
1 0.83\\
2    0.83\\
3    0.89\\
4    1.07\\
};
\addlegendentry{RandomK}

\addplot[ybar, fill=violmp, area legend, postaction={pattern=crosshatch dots}] table[row sep=crcr] {%
1 5.7242 \\ 
2 5.27165 \\
3 3.82955 \\
4 1.4277 \\
};
\addlegendentry{GBS}

\end{axis}

\end{tikzpicture}
		\caption{Uniformly aperiodic traffic.}
		\label{fig:comparison_traffic_unif}
	\end{subfigure}\\\vspace{0.33cm}
	\begin{subfigure}[t!]{0.99\columnwidth}
		\centering 
		\setlength\fwidth{0.8\columnwidth}
		\setlength\fheight{0.25\columnwidth}
\pgfplotsset{scaled y ticks=false}
\pgfplotsset{scaled x ticks=false}
\begin{tikzpicture}
\pgfplotsset{
tick label style={font=\scriptsize},
label style={font=\scriptsize},
legend  style={font=\scriptsize}
}

\definecolor{darkgray176}{RGB}{176,176,176}
\definecolor{bluemp}{RGB}{2,191,255}
\definecolor{redmp}{RGB}{205,92,91}
\definecolor{violmp}{RGB}{230,230,250}
\definecolor{orangep}{RGB}{255,159,122,255}

\begin{axis}[
width=0.951\fwidth,
height=\fheight,
at={(0\fwidth,0\fheight)},
scale only axis,
 every axis plot post/.style={/pgf/number format/fixed},
    ybar=5pt,
    bar width=8pt,
    xmin=0.5,
	xmax=4.5,
	xtick={1,2,3,4},
	xticklabels={1,2,3,5},
    xlabel={$t_{min}$ [ms]},
	xlabel style={font=\scriptsize\color{white!15!black}},
	xticklabel style={align=center},
    ymin=0,
    axis on top,
    ymax=2,
    xmajorgrids,
    ymajorgrids,
    yminorticks=true,
	ylabel style={font=\scriptsize\color{white!15!black}},
	ylabel={\gls{e2e} latency [ms]},
    restrict y to domain*=0:2.5, 
    visualization depends on=rawy\as\rawy, 
    after end axis/.code={ 
            \draw [ultra thick, white, decoration={snake, amplitude=1pt}, decorate] (rel axis cs:0,1.08) -- (rel axis cs:1,1.08);
        },
    nodes near coords={%
           \tiny\pgfmathprintnumber{\rawy}
        },
    axis lines*=left,
    clip=false,
    ]
\addplot[ybar, fill=bluemp, area legend, postaction={pattern=north east lines, opacity=0.5}] table[row sep=crcr] {%
1 0.71    \\ 
2 0.71   \\ 
3 0.69   \\ 
4 0.72\\ 
};

\addplot[ybar, fill=redmp, area legend] table[row sep=crcr] {%
1 0.89\\
2     0.87\\
3     0.90\\
4    1.15\\
};

\addplot[ybar, fill=violmp, area legend, postaction={pattern=crosshatch dots}] table[row sep=crcr] {%
1 4.19935\\
2 3.7495\\
3 2.76295 \\
4 1.3634 \\
};

\end{axis}

\end{tikzpicture}
		\caption{UE-specific aperiodic traffic.}
		\label{fig:comparison_traffic_ue}
	\end{subfigure}
	\caption{Average \gls{e2e} latency for \gls{disnets}, RandomK, and GBS, as a function of $t_{min}$ and the type of traffic. We set $t_{max}=6$ ms.\vspace{-0.33cm}}
	\label{fig:perf_traffic}
\end{figure}
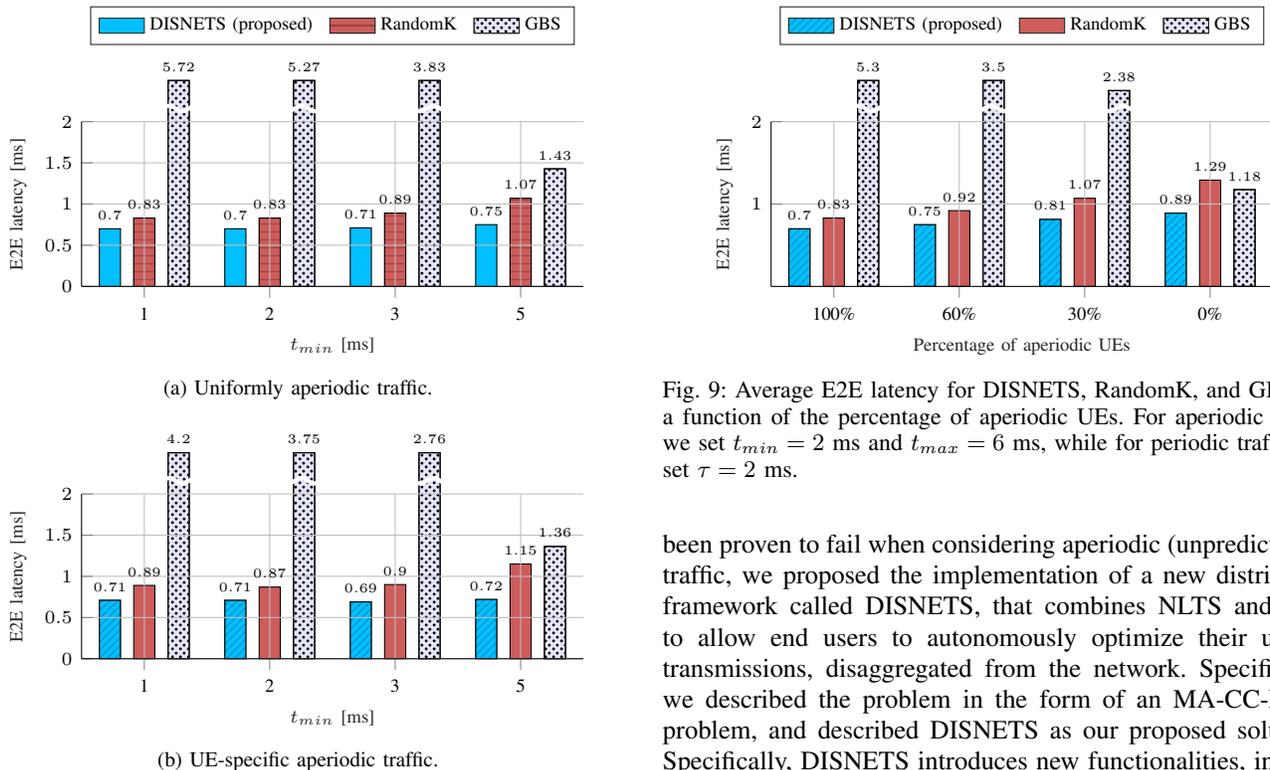

From the previous paragraphs we concluded that \gls{sps} and \gls{nlts} are not compatible with URLLC requirements for aperiodic traffic.
So, in this set of experiments we focus on \gls{disnets}, RandomK, and GBS as a function of the type of traffic.
First, in Fig.~\ref{fig:perf_traffic} we change the value of $t_{min}$, which is inversely proportional to the traffic load, considering both uniformly aperiodic (Fig.~\ref{fig:comparison_traffic_unif}) and UE-specific aperiodic traffic (Fig.~\ref{fig:comparison_traffic_ue}).
We can see that \gls{disnets} is better than any other benchmark, and the latency is consistently below 1 ms in all configurations.
Notice that for GBS the \gls{e2e} latency increases as $t_{min}$ decreases because the traffic is more intense and the system is more congested. 
On the contrary, for RandomK and \gls{disnets}  we have the opposite trend, i.e., the \gls{e2e} latency increases as $t_{min}$ increases.
This is motivated by the fact that, as $t_{min}$ approaches $t_{max}=6$ ms, the traffic becomes quasi-deterministic and the \glspl{ue} tend to generate packets almost simultaneously, which may increase the number of collisions. Consequently, achieving coordination becomes harder.
Still, for \gls{disnets} the latency grows as little as 7\%, from 0.7 for $t_{min}=1$ ms to 0.75 ms for $t_{min}=5$ ms.
In turn, the performance of RandomK deteriorates significantly as $t_{min}$ increases, and almost achieves the same performance as GBS in the long term.


\begin{figure}[!t]
	\centering 
	\begin{subfigure}[t!]{0.99\columnwidth}
		\centering 
		\setlength\fwidth{0.8\columnwidth}
		\setlength\fheight{0.25\columnwidth}
\pgfplotsset{scaled y ticks=false}
\pgfplotsset{scaled x ticks=false}
\begin{tikzpicture}
\pgfplotsset{
tick label style={font=\scriptsize},
label style={font=\scriptsize},
legend  style={font=\scriptsize}
}

\definecolor{darkgray176}{RGB}{176,176,176}
\definecolor{bluemp}{RGB}{2,191,255}
\definecolor{redmp}{RGB}{205,92,91}
\definecolor{violmp}{RGB}{230,230,250}
\definecolor{orangep}{RGB}{255,159,122,255}

\begin{axis}[
width=0.951\fwidth,
height=\fheight,
at={(0\fwidth,0\fheight)},
scale only axis,
 every axis plot post/.style={/pgf/number format/fixed},
    ybar=5pt, 
    bar width=8pt,
    xmin=0.5,
	xmax=4.5,
	xtick={1,2,3,4},
	xticklabels={100\%,   60\%,  30\%,  0\%},
	xlabel={Percentage of aperiodic UEs},
	xlabel style={font=\scriptsize\color{white!15!black}},
	xticklabel style={align=center},
    ymin=0,
    axis on top,
    ymax=2,
    xmajorgrids,
    ymajorgrids,
    yminorticks=true,
    ytick={1,2,3,4,5},
	ylabel style={font=\scriptsize\color{white!15!black}},
	ylabel={\gls{e2e} latency [ms]},
    restrict y to domain*=0:2.5, 
    visualization depends on=rawy\as\rawy, 
    after end axis/.code={ 
            \draw [ultra thick, white, decoration={snake, amplitude=1pt}, decorate] (rel axis cs:0,1.08) -- (rel axis cs:1,1.08);
      },
    nodes near coords={%
           \tiny\pgfmathprintnumber{\rawy}
        },
    axis lines*=left,
    clip=false,
    legend style={legend cell align=left, align=left, draw=white!15!black, at={(0.5,1.7)},/tikz/every even column/.append style={column sep=0.15cm},
  anchor=north ,legend columns=-1},
    ]
\addplot[ybar, fill=bluemp, area legend, postaction={pattern=north east lines, opacity=0.5}] table[row sep=crcr] {%
1 0.70    \\
2 0.75\\
3    0.8144\\
4    0.89\\
};
\addlegendentry{\gls{disnets} (proposed)}

\addplot[ybar, fill=redmp, area legend] table[row sep=crcr] {%
1 0.83\\
2    0.92\\
3    1.07\\
4    1.29\\
};
\addlegendentry{RandomK}

\addplot[ybar, fill=violmp, area legend, postaction={pattern=crosshatch dots}] table[row sep=crcr] {%
1 5.3\\
2 3.5\\
3 2.37785\\
4 1.1753\\
};
\addlegendentry{GBS}

\end{axis}

\end{tikzpicture}
	\end{subfigure}
	\caption{Average \gls{e2e} latency for \gls{disnets}, RandomK, and GBS, as a function of the percentage of aperiodic UEs. For aperiodic traffic we set $t_{min}=2$ ms and $t_{max}=6$ ms, while for periodic traffic we set $\tau = 2$ ms.\vspace{-0.33cm}}
	\label{fig:comparison_perc}
\end{figure}
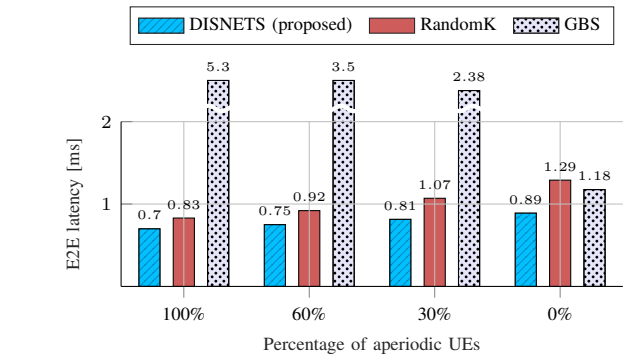

Finally, in Fig.~\ref{fig:comparison_perc} we study the \gls{e2e} latency as a function of the percentage of aperiodic \glspl{ue} in the network. 
Specifically, for the fraction of aperiodic \glspl{ue} we set $t_{min} = 2$ ms and $t_{max} = 6$ ms, whereas for the periodic UEs we set $\tau = 2$ ms. 
As such, the average inter-packet interval for aperiodic UEs is equal to $4$ ms vs. $2$ ms for periodic UEs, which means that the latter generates more traffic.
Again, we see that \gls{disnets} outperforms the other benchmarks, and is therefore able to work well in both periodic and mixed, i.e., periodic and aperiodic, traffic conditions.
Notably, the performance of GBS decreases as the traffic becomes more aperiodic. 
This is expected, and we proved in~\cite{cuozzo2022enabling,cavallero2023new} that GBS, as well as \gls{sps}, do not work well in case of unpredictable aperiodic traffic. 
For example, in case of GBS, some packets may be generated towards the end of an activation period, and cannot send scheduling requests within the current activation period, so packets remain in the queue and accumulate delay.
On the other hand, as mentioned in Fig.~\ref{fig:perf_traffic}, \gls{disnets} and RandomK have the opposite trend, and suffer more when the traffic becomes periodic given that periodic UEs generate more traffic than aperiodic UEs.
Still, \gls{disnets} is able to converge to good and stable results, and decrease the latency by up to $86\%$ and $50\%$  compared to {GBS} and RandomK, respectively.

\paragraph{DISNETS vs. RandomK}
The above results demonstrate the superior performance of DISNETS compared to RandomK under several metrics. 
For example, for 100 UEs, DISNETS can reduce the latency by up to 20\% (see Fig.~\ref{fig:perf_ue}), and improve reliability by up to 40\% (see Fig.~\ref{fig:perf_traffic}).
Moreover,  RandomK requires offline simulations to identify the optimal value of $K$, i.e., $K^*$ (since $K^*$ changes as changing the traffic periodicity and the number of UEs), which may be time consuming and not always feasible in practice. On the other hand, DISNETS is flexible enough to adapt to different scenarios after the training phase (which is done only once).

\section{Conclusion and Future Work}
\label{sec:conclusion}

In this paper we shed light on the issue of enabling \gls{urllc} in \gls{iiot} networks. Specifically, we focused on the impact of resource allocation on the \gls{e2e} latency. While the two main 5G NR centralized schedulers, namely GBS and SPS, have been proven to fail when considering aperiodic (unpredictable) traffic, we proposed the implementation of a new distributed framework called \gls{disnets}, that combines \gls{nlts} and \gls{lts} to allow end users to autonomously optimize their uplink transmissions, disaggregated from the network.
Specifically, we described the problem in the form of an \gls{maccmab} problem, and described \gls{disnets} as our proposed solution. 
Specifically, \gls{disnets} introduces new functionalities, including (i) a new control signaling scheme called \gls{fci} to train the \gls{disnets} framework, and (ii) a new protocol procedure for autonomously selecting multiple radio resources to reduce the probability of collision.
We showed via simulations that \gls{disnets} is compatible with URLLC even for aperiodic traffic and considering IIoT-specific correlations, and outperforms state-of-the-art centralized and decentralized benchmarks.

As part of our future work, we plan to develop a scalable and practical demonstrator using 5G off-the-shelf commercial equipement to emulate distributed resource allocation for IIoT URLLC based on the studies of this paper.

\bibliographystyle{IEEEtran}
\bibliography{bibl.bib}

\end{document}